\documentclass[a4paper]{IEEEtran}

\usepackage{color}

\ifCLASSINFOpdf
\usepackage[pdftex]{graphicx}
\else
\usepackage[dvips]{graphicx}
\DeclareGraphicsExtensions{.eps}
\fi
\usepackage{epstopdf}

\usepackage{graphicx}
\usepackage{caption}
\usepackage{subcaption}

\usepackage{amsmath, amsthm, amssymb} % odkazy na rovnice
\usepackage{upgreek} % psat rekce mu vzprimene

\usepackage{algorithm} % prostredi algorithm
\usepackage{algpseudocode} % pro algorithm
\floatstyle{plain}
\newfloat{myalgo}{tbhp}{mya}
{\begin{myalgo}[#1]
\centering
\begin{minipage}{#2}
\begin{algorithm}[H]}%
{\end{algorithm}
\end{minipage}
\end{myalgo}}
\usepackage{setspace}

\usepackage{enumitem}

\begin{document}
\title{Mobile Edge Computing: A Survey on Architecture and Computation Offloading}

\author{\IEEEauthorblockN{Pavel Mach, \emph{IEEE Member}, Zdenek Becvar, \emph{IEEE Member}} \\
\thanks{This work has been supported by the grant of Czech Technical University in Prague No. SGS17/184/OHK3/3T/13.}
\thanks{The authors are with the Department of Telecommunication Engineering, Faculty of Electrical Engineering, Czech Technical University in Prague, Prague, 166 27 Czech Republic (email:machp2@fel.cvut.cz; zdenek.becvar@fel.cvut.cz).}
}

\maketitle

\begin{abstract}
Technological evolution of mobile user equipments (UEs), such as smartphones or laptops, goes hand-in-hand with evolution of new mobile applications. However, running computationally demanding applications at the UEs is constrained by limited battery capacity and energy consumption of the UEs. Suitable solution extending the battery life-time of the UEs is to offload the applications demanding huge processing to a conventional centralized cloud (CC). Nevertheless, this option introduces significant execution delay consisting in delivery of the offloaded applications to the cloud and back plus time of the computation at the cloud. Such delay is inconvenient and make the offloading unsuitable for real-time applications. To cope with the delay problem, a new emerging concept, known as mobile edge computing (MEC), has been introduced. The MEC brings computation and storage resources to the edge of mobile network enabling to run the highly demanding applications at the UE while meeting strict delay requirements. The MEC computing resources can be exploited also by operators and third parties for specific purposes. \textcolor{black}{In this paper, we first describe major use cases and reference scenarios where the MEC is applicable. After that we survey existing concepts integrating MEC functionalities to the mobile networks and discuss current advancement in standardization of the MEC.} The core of this survey is, then, focused on user-oriented use case in the MEC, i.e., computation offloading. In this regard, we divide the research on computation offloading to three key areas: i) decision on computation offloading, ii) allocation of computing resource within the MEC, and iii) mobility management. \textcolor{black}{Finally, we highlight lessons learned in area of the MEC and we discuss open research challenges yet to be addressed in order to fully enjoy potentials offered by the MEC.}
\end{abstract}

%\begin{IEEEkeywords}
 
%\end{IEEEkeywords}

% \fi
\IEEEpeerreviewmaketitle

\section{Introduction}
The users' requirements on data rates and quality of service (QoS) are exponentially increasing. Moreover, technological evolution of smartphones, laptops and tablets enables to emerge new high demanding services and applications. Although new mobile devices are more and more powerful in terms of central processing unit (CPU), even these may not be able to handle the applications requiring huge processing in a short time. Moreover, high battery consumption still poses a significant obstacle restricting the users to fully enjoy highly demanding applications on their own devices. This motivates development of mobile cloud computing (MCC) concept allowing cloud computing for mobile users \cite{Hoang2013}. In the MCC, a user equipment (UE) may exploit computing and storage resources of powerful distant centralized clouds (CC), which are accessible through a core network (CN) of a mobile operator and the Internet. The MCC brings several advantages \cite{Barbarossa2014}; 1) extending battery lifetime by offloading energy consuming computations of the applications to the cloud, 2) enabling sophisticated applications to the mobile users, and 3) providing higher data storage capabilities to the users. Nevertheless, the MCC also imposes huge additional load both on radio and backhaul of mobile networks and introduces high latency since data is sent to powerful farm of servers that are, in terms of network topology, far away from the users. 

To address the problem of a long latency, the cloud services should be moved to a proximity of the UEs, i.e., to the edge of mobile network as considered in newly emerged edge computing paradigm. \textcolor{black}{The edge computing can be understood as a specific case of the MCC. Nevertheless, in the conventional MCC, the cloud services are accessed via the Internet connection \cite{Khan2014} while in the case of the edge computing, the computing/storage resources are supposed to be in proximity of the UEs (in sense of network topology).} Hence, the MEC can offer significantly lower latencies and jitter when compared to the MCC. Moreover, while the MCC is fully centralized approach with farms of computers usually placed at one or few locations, the edge computing is supposed to be deployed in fully distributed manner. On the other hand, the edge computing provides only limited computational and storage resources with respect to the MCC. A high level comparison of key technical aspects of the MCC and the edge computing is outlined in Table~\ref{tab:Tab1}.

\begin{table}[b!]
\footnotesize
\caption{High level comparison of MCC and Edge computing concepts}
\label{tab:Tab1}
\centering
\renewcommand{\arraystretch}{1.2}
\begin{tabular}{|p{2.5cm}||p{2cm}|p{2cm}| }
\hline

	\bf Technical aspect & \bf MCC & \bf Edge computing \\ \hline \hline
	Deployment & Centralized & Distributed \\ \hline
	Distance to the UE & High & Low \\ \hline
	Latency & High & Low \\ \hline
	Jitter & High & Low \\ \hline
	Computational power  & Ample & Limited  \\ \hline
	Storage capacity & Ample & Limited \\ \hline
\end{tabular}
\end{table}

The first edge computing concept bringing the computation/storage closer to the UEs, proposed in 2009, is cloudlet \cite{Satyanarayanan2009}. The idea behind the cloudlet is to place computers with high computation power at strategic locations in order to provide both computation resources and storage for the UEs in vicinity. The cloudlet concept of the computing "hotspots" is similar to WiFi hotspots scenario, but instead of Internet connectivity the cloudlet enables cloud services to the mobile users. The fact that cloudlets are supposed to be mostly accessed by the mobile UEs through WiFi connection is seen as a disadvantage since the UEs have to switch between the mobile network and WiFi whenever the cloudlet services are exploited \cite{Barbarossa2014}. Moreover, QoS (Quality of Service) of the mobile UEs is hard to fulfill similarly as in case of the MCC, since the cloudlets are not an inherent part of the mobile network and coverage of WiFi is only local with limited support of mobility.

The other option enabling cloud computing at the edge is to perform computing directly at the UEs through ad-hoc cloud allowing several UEs in proximity to combine their computation power and, thus, process high demanding applications locally \cite{Shi2012}-\textcolor{black}{\cite{Zhang2013a}}. To facilitate the ad-hoc cloud, several critical challenges need to be addressed; 1) finding proper computing UEs in proximity while guaranteeing that processed data will be delivered back to the source UE, 2) coordination among the computing UEs has to be enabled despite the fact that there are no control channels to facilitate reliable computing, 3) the computing UEs has to be motivated to provide their computing power to other devices given the battery consumption and additional data transmission constraints, 4) security and privacy issues.

A more general concept of the edge computing, when compared to cloudlets and ad-hoc clouds, is known as a fog computing. The fog computing paradigm (shortly often abbreviated as Fog in literature) has been introduced in 2012 by Cisco to enable a processing of the applications on billions of connected devices at the edge of network \cite{Bonomi2012}. Consequently, the fog computing may be considered as one of key enablers of Internet of Things (IoT) and big data applications \cite{Zhu2013} as it offers: 1) low latency and location awareness due to proximity of the computing devices to the edge of the network, 2) wide-spread geographical distribution when compared to the CC; 3) interconnection of very large number of nodes (e.g., wireless sensors), and 4) support of streaming and real time applications \cite{Bonomi2012}. Moreover, the characteristics of the fog computing can be exploited in many other applications and scenarios such as smart grids, connected vehicles for Intelligent Transport Systems (ITS) or wireless sensor networks \cite{Stojmenovic2014}-\cite{Yannuzzi2014}.

\textcolor{black}{From the mobile users' point of view, the most notable drawback of all above-mentioned edge computing concepts is that QoS and QoE (Quality of Experience) for users can be hardly guaranteed, since the computing is not integrated into an architecture of the mobile network. One concept integrating the cloud capabilities into the mobile network is Cloud Radio Access Network (C-RAN) \cite{Checko2015}. \textcolor{black}{The C-RAN exploits the idea of distributed protocol stack \cite{Kliazovich2008}, where some layers of the protocol stack are moved from distributed Radio Remote Heads (RRHs) to centralized baseband units (BBUs)}. The BBU's computation power is, then, pooled together into virtualized resources that are able to serve tens, hundreds or even thousands of RRHs. Although the computation power of this virtualized BBU pool is exploited primarily for a centralized control and baseband processing it may also be used for the computation offloading to the edge of the network (see, for example, \cite{Cheng2016}).}

\textcolor{black}{Another concept integrating the edge computing into the mobile network architecture is developed by newly created (2014) industry specification group (ISG) within European Telecommunications Standards Institute (ETSI) \cite{Hu2015}. The solution outlined by ETSI is known as Mobile Edge Computing (MEC). The standardization efforts relating the MEC are driven by prominent mobile operators (e.g., DOCOMO, Vodafone, TELECOM Italia) and manufactures (e.g., IBM, Nokia, Huawei, Intel). The main purpose of ISG MEC group is to enable an efficient and seamless integration of the cloud computing functionalities into the mobile network, and to help developing favorable conditions for all stakeholders (mobile operators, service providers, vendors, and users).} 

\textcolor{black}{Several surveys on cloud computing have been published so far. In \cite{Khan2014}, the authors survey MCC application models and highlight their advantages and shortcomings. In \cite{Sanaei2014}, a problem of a heterogeneity in the MCC is tackled. The heterogeneity is understood as a variability of mobile devices, different cloud vendors providing different services, infrastructures, platforms, and various communication medium and technologies. The paper identifies how this heterogeneity impacts the MCC and discusses related challenges. The authors in \cite{Wen2014} survey existing efforts on Cloud Mobile Media, which provides rich multimedia services over the Internet and mobile wireless networks. All above-mentioned papers focus, in general, on the MCC where the cloud is not allocated specifically at the edge of mobile network, but it is accessed through the Internet. Due to a wide potential of the MEC, there is a lot of effort both in industry and academia focusing on the MEC in particular. Despite this fact, there is just one survey focusing primarily on the MEC \cite{Ahmed2016} that, however, only briefly surveys several research works dealing with the MEC and presents taxonomy of the MEC by describing key attributes. Furthermore, the authors in \cite{Roman2016} extensively surveys security issues for various edge computing concepts. On top of that, the authors in \cite{Luong2017} dedicate one chapter to the edge computing, where applications of economic and pricing models are considered for resource management in the edge computing.} 

\begin{figure*}[t!]
	\centering
	\includegraphics[scale=0.088]{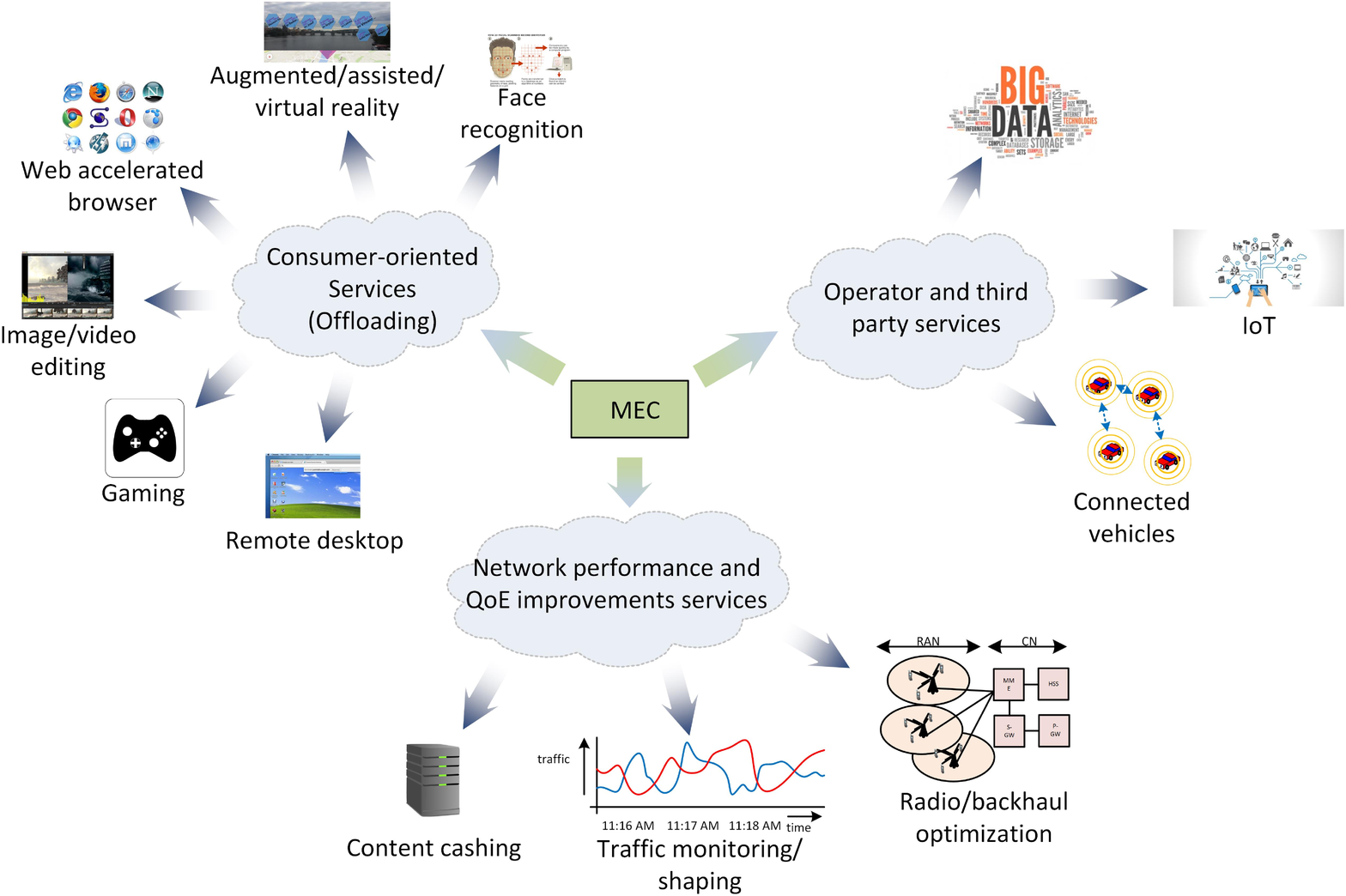} 
	\caption{\textcolor{black}{Example of use cases and scenarios for the MEC.}}
	\label{fig:new}
\end{figure*}

\textcolor{black}{In contrast to the above-mentioned surveys, we describe key use cases and scenarios for the MEC (Section~\ref{Usecases}). Then, we survey existing MEC concepts proposed in the literature integrating the MEC functionalities into the mobile networks and we discuss standardization  of the MEC (Section~\ref{sec2}). After that, the core part of the paper is focused on technical works dealing with computation offloading to the MEC. On one hand, the computation offloading can be seen as a key use case from the user perspective as it enables running new sophisticated applications at the UE while reducing its energy consumption (see, e.g., \cite{MAUI}-\cite{Flores2015} where computation offloading to distant CC is assumed). On the other hand, the computation offloading brings several challenges, such as selection of proper application and programming models, accurate estimation of energy consumption, efficient management of simultaneous offloading by multiple users, or virtual machine (VM) migration \cite{Jiao2013}. In this respect, we overview several general principles related to the computation offloading, such as offloading classification (full, partial offloading), factors influencing the offloading itself, and management of the offloading in practice (Section~\ref{offloading}). Afterwards, we sort the efforts within research community addressing following key challenges regarding computation offloading into the MEC:}

\begin{itemize}
\item 
A \textbf{decision on the computation offloading} to the MEC with the purpose to determine whether the offloading is profitable for the UE in terms of energy consumption and/or execution delay (Section~\ref{sec3}).
\item 
An efficient \textbf{allocation of the computing resources} within the MEC if the computation is offloaded in order to minimize execution delay and balance load of both computing resources and communication links (Section~\ref{sec4}).
\item
\textbf{Mobility management} for the applications offloaded to the MEC guaranteeing service continuity if the UEs exploiting the MEC roams throughout the network  (Section~\ref{sec5}).
\end{itemize}
Moreover, \textcolor{black}{we summarize the lessons learned from state of the art focused on computation offloading to the MEC (Section VIII)} and outline several open challenges, which need to be addressed to make the MEC beneficial for all stakeholders (Section~\ref{sec6}). Finally, we summarize general outcomes and draw conclusions (Section~\ref{sec7}).

\section{\textcolor{black}{Use cases and service scenarios}}
\label{Usecases}
\textcolor{black}{The MEC brings many advantages to all stakeholders, such as mobile operators, service providers or users. As suggested in \cite{MEC002}\cite{Beck2014}, three main use case categories, depending on the subject to which they are profitable to, can be distinguished for the MEC (see Fig.~\ref{fig:new}). The next subsections discuss individual use case categories and pinpoint several key service scenarios and applications.}

\subsection{\textcolor{black}{Consumer-oriented services}}
\label{Usecases1}
\textcolor{black}{The first use case category is consumer-oriented and, hence, should be beneficial directly to the end-users. In general, the users profit from the MEC mainly by means of the computation offloading, which enables running new emerging applications at the UEs. One of the applications benefiting from the computation offloading is a web accelerated browser, where most of the browsing functions (web contents evaluation, optimized transmission, etc.) are offloaded to the MEC; see experimental results on offloading of web accelerated browser to the MEC in \cite{Taka2015}. Moreover, face/speech recognition or image/video editing are also suitable for the MEC as these require large amount of computation and storage \cite{Zhang2012}.} 

\textcolor{black}{Besides, the computation offloading to the MEC can be exploited by the applications based on augmented, assisted or virtual reality. These applications derive additional information about users' neighborhood by performing an analysis of their surroundings (e.g., visiting tourists may find points of interest in his/her proximity). This may require fast responses, and/or significant amount of computing resources not available at the UE. An applicability of the MEC for augmented reality is shown in \cite{Dolezal2016}. The authors demonstrate on a real MEC testbed that the reduction of latency up to 88\% and energy consumption of the UE up to 93\% can be accomplished by the computation offloading to the MEC.} 

\textcolor{black}{On top of that, the users running low latency applications, such as online gaming or remote desktop, may profit from the MEC in proximity. In this case a new instance of a specific application is initiated at an appropriate mobile edge host to reduce the latency and resources requirements of the application at the UE.}

\subsection{\textcolor{black}{Operator and third party services}}
\label{Usecases2}
\textcolor{black}{The second use case category is represented by the services from which operators and third parties can benefit. An example of the use case profitable for the operator or third party is a gathering of a huge amount of data from the users or sensors. Such data is first pre-processed and analyzed at the MEC. The pre-processed data is, then, sent to distant central servers for further analysis. This could be exploited for safety and security purposes, such as monitoring of an area (e.g., car park monitoring).}

\textcolor{black}{Another use case is to exploit the MEC for IoT (Internet of Thing) purposes \cite{Salman2015}-\cite{Sun2016a}. Basically, IoT devices are connected through various radio technologies (e.g., 3G, LTE, WiFi, etc.) using diverse communication protocols. Hence, there is a need for low latency aggregation point to handle various protocols, distribution of messages and for processing. This can be enabled by the MEC acting as an IoT gateway, which purpose is to aggregate and deliver IoT services into highly distributed mobile base stations in order to enable applications responding in real time.}  

\textcolor{black}{The MEC can be also exploited for ITS to extend the connected car cloud into the mobile network. Hence, roadside applications running directly at the MEC can receive local messages directly from applications in the vehicles and roadside sensors, analyze them and broadcast warnings (e.g., an accident) to nearby vehicles with very low latency. The exploitation of the MEC for car-to-car and car-to-infrastructure communications was demonstrated by Nokia and its partners in an operator's LTE network just recently in 2016 \cite{Nokia2016}\cite{Nokia2016a}.} 

\subsection{\textcolor{black}{Network performance and QoE improvement services}}
\label{Usecases3}
\textcolor{black}{The third category of use cases are those optimizing network performance and/or improving QoE. One such use case is to enable coordination between radio and backhaul networks. So far, if the capacity of either backhaul or radio link is degraded, the overall network performance is negatively influenced as well, since the other part of the network (either radio or backhaul, respectively) is not aware of the degradation. In this respect, an analytic application exploiting the MEC can provide real-time information on traffic requirements of the radio/backhaul network. Then, an optimization application, running on the MEC, reshapes the traffic per application or re-routes traffic as required.}

\textcolor{black}{Another way to improve performance of the network is to alleviate congested backhaul links by local content caching at the mobile edge. This way, the MEC application can store the most popular content used in its geographical area. If the content is requested by the users, it does not have to be transfered over the backhaul network. }
	
\textcolor{black}{Besides alleviation and optimization of the backhaul network, the MEC can also help in radio network optimization. For example, gathering related information from the UEs and processing these at the edge will result in more efficient scheduling. In addition, the MEC can also be used for mobile video delivery optimization using throughput guidance for TCP (Transmission Control Protocols). The TCP has an inherent difficulty to adapt to rapidly varying condition on radio channel resulting in an inefficient use of the resources. To deal with this problem, the analytic MEC application can provide a real-time indication on an estimated throughput to a backend video server in order to match the application-level coding to the estimated throughput.}     

\begin{figure*}[t!]
	\centering
	\includegraphics[scale=0.09]{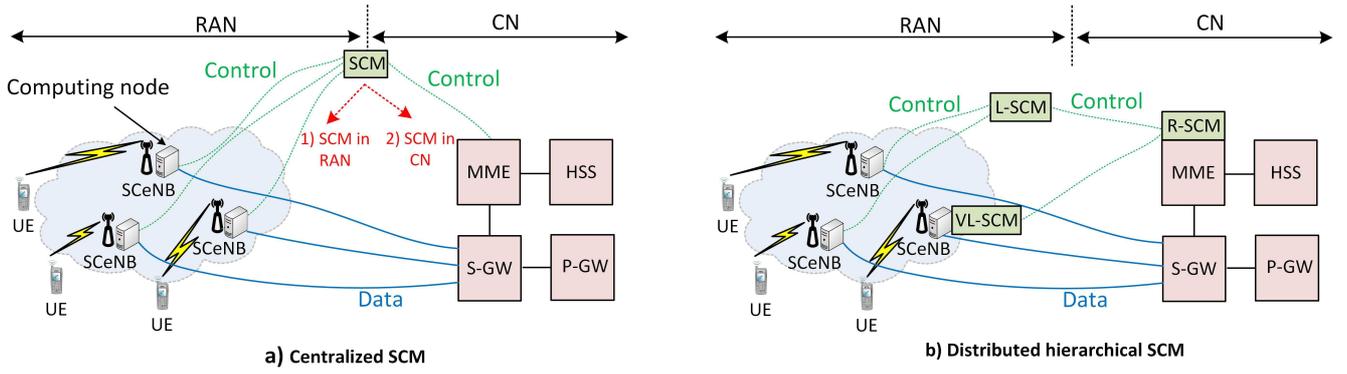} 
	\caption{SCC architecture (MME - Mobility Management Entity, HSS - Home Subscriber Server, S-GW - Serving Gateway, P-GW - Packet Gateway).}
	\label{fig:01}
\end{figure*}

\section{MEC Architecture and standardization}
\label{sec2}
This section introduces and compares several concepts for the computation at the edge integrated to the mobile network. First, we overview various MEC solutions proposed in the literature that enable to bring computation close to the UEs. Secondly, we describe the effort done within ETSI standardization organization regarding the MEC. Finally, we compare individual existing MEC concepts (proposed in both literature and ETSI) from several perspectives, such as MEC control or location of the computation/storage resources.

\subsection{Overview of the MEC concept}
\label{sec21}
In recent years, several MEC concepts with purpose to smoothly integrate cloud capabilities into the mobile network architecture have been proposed in the literature. This section briefly introduces fundamental principles of small cell cloud (SCC), mobile micro cloud (MMC), fast moving personal cloud, follow me cloud (FMC), and CONCERT. Moreover, the section shows enhancements/modifications to the network architecture necessary for implementation of each MEC concept.

\subsubsection{Small cell cloud (SCC)}
\label{sec211}

The basic idea of the SCC, firstly introduced in 2012 by the European project TROPIC \cite{TROPIC2012}\cite{Lobillo2014}, is to enhance small cells (SCeNBs), like microcells, picocells or femtocells, by an additional computation and storage capabilities. The similar idea is later on addressed in SESAME project as well, where the cloud-enabled SCeNBs supports the edge computing \cite{SESAM2015}\cite{Giannoulakis2016}. The cloud-enhanced SCeNBs can pool their computation power exploiting network function virtualization (NFV) \cite{NFV2012}\cite{NFV2013} paradigm. Because a high number of the SCeNBs is supposed to be deployed in future mobile networks, the SCC can provide enough computation power for the UEs, especially for services/applications having stringent requirements on latency (the examples of such applications are listed in Section~\ref{Usecases1}). 

In order to fully and smoothly integrate the SCC concept into the mobile network architecture, a new entity, denoted as a small cell manager (SCM), is introduced to control the SCC \cite{Lobillo2014}. The SCM is in charge of the management of the computing and/or storage resources provided by the SCeNBs. Since the SCeNBs can be switched on/off at any time (especially if owned by the users as in case of the femtocells), the SCM performs dynamic and elastic management of the computation resources within the SCC. The SCM is aware of the overall cluster context (both radio and cloud-wise) and decides where to deploy a new computation or when to migrate an on-going computation to optimize the service delivery for the end-user. The computing resources are virtualized by means of Virtual Machine (VM) located at the SCeNBs. An important aspect regarding the architecture of the SCC is deployment of the SCM (see Fig.~\ref{fig:01}). The SCM may be deployed in a centralized manner either as a standalone SCM located within the RAN, close to a cluster of the SCeNBs, or as an extension to a MME \cite{Lobillo2014}\cite{Puente2015}. Moreover, the SCM can be deployed also in a distributed hierarchical manner, where a local SCM (L-SCM) or a virtual L-SCM (VL-SCM) manages the computing and storage resources of the SCeNBs' clusters in vicinity while a remote SCM (R-SCM), located in the CN, has resources of all SCeNBs connected to the CN at its disposal \cite{Becvar2017} (see Fig.~\ref{fig:01}b).

\begin{figure}[b!]
	\centering
	\includegraphics[scale=0.078]{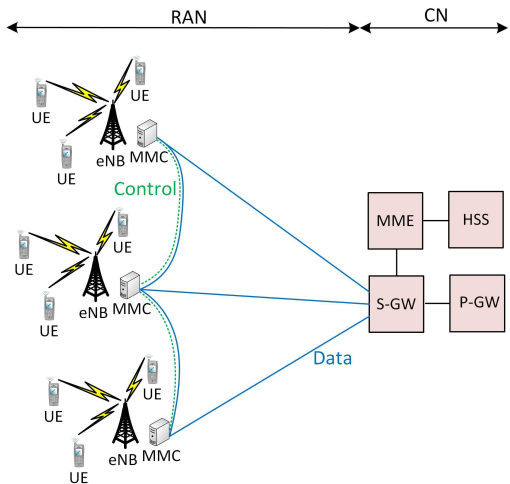} 
	\caption{MMC architecture.}
	\label{fig:02}
\end{figure}

\subsubsection{Mobile micro clouds (MMC)}
\label{sec212}
The concept of the MMC has been firstly introduced in \cite{Wang2013}. Like the SCC, also the MMC allows users to have instantaneous access to the cloud services with a low latency. While in the SCC the computation/storage resources are provided by interworking cluster(s) of the SCeNBs, the UEs exploit the computation resources of a single MMC, which is typically connected directly to a wireless base station (i.e., the eNB in the mobile network) as indicated in Fig.~\ref{fig:02}. The MMC concept does not introduce any control entity into the network and the control is assumed to be fully distributed in a similar way as the VL-SCM solution for the SCC. To this end, the MMCs are interconnected directly or through backhaul in order to guarantee service continuity if the UEs move within the network to enable smooth VM migration among the MMCs (see more detail on VM migration in Section~\ref{sec52}).

 \begin{figure}[b!]
	\centering
	\includegraphics[scale=0.075]{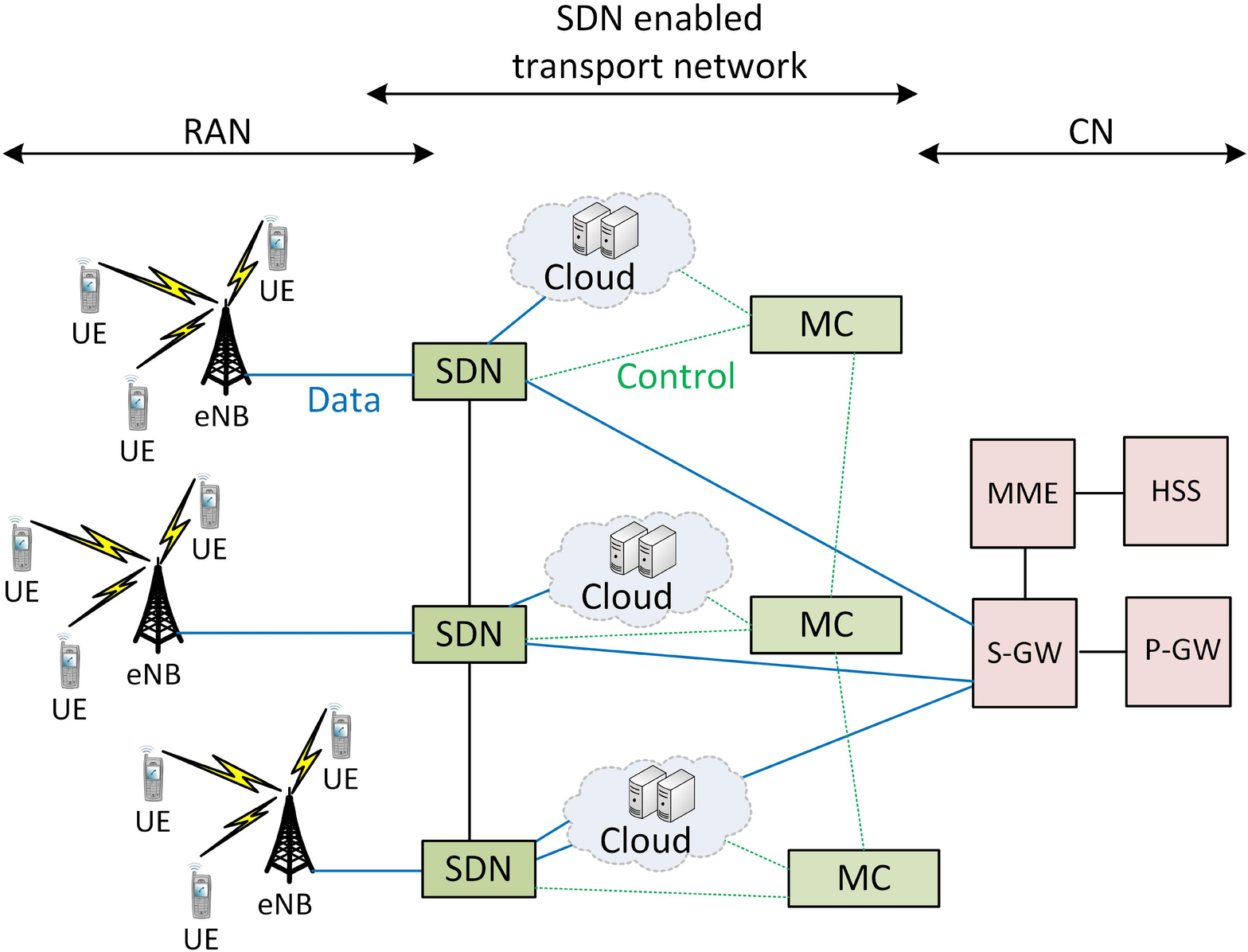} 
	\caption{MobiScud architecture \cite{Wang2015}.}
	\label{fig:03}
\end{figure}

\subsubsection{Fast moving personal cloud (MobiScud)}
\label{sec213}
The MobiScud architecture \cite{Wang2015} integrates the cloud services into the mobile networks by means of software defined network (SDN) \cite{Manzalini2014} and NFV technologies whilst maintaining backward compatibility with existing mobile network. When compared to the SCC and the MMC concepts, the cloud resources in the MobiScud are not located directly at the access nodes such as SCeNB or eNB, but at operator's  clouds located within RAN or close to RAN (see Fig.~\ref{fig:03}). Still, these clouds are assumed to be highly distributed similarly as in case of the SCC and the MMC enabling the cloud service to all UEs in vicinity. 

\begin{figure}[b!]
	\centering
	\includegraphics[scale=0.07]{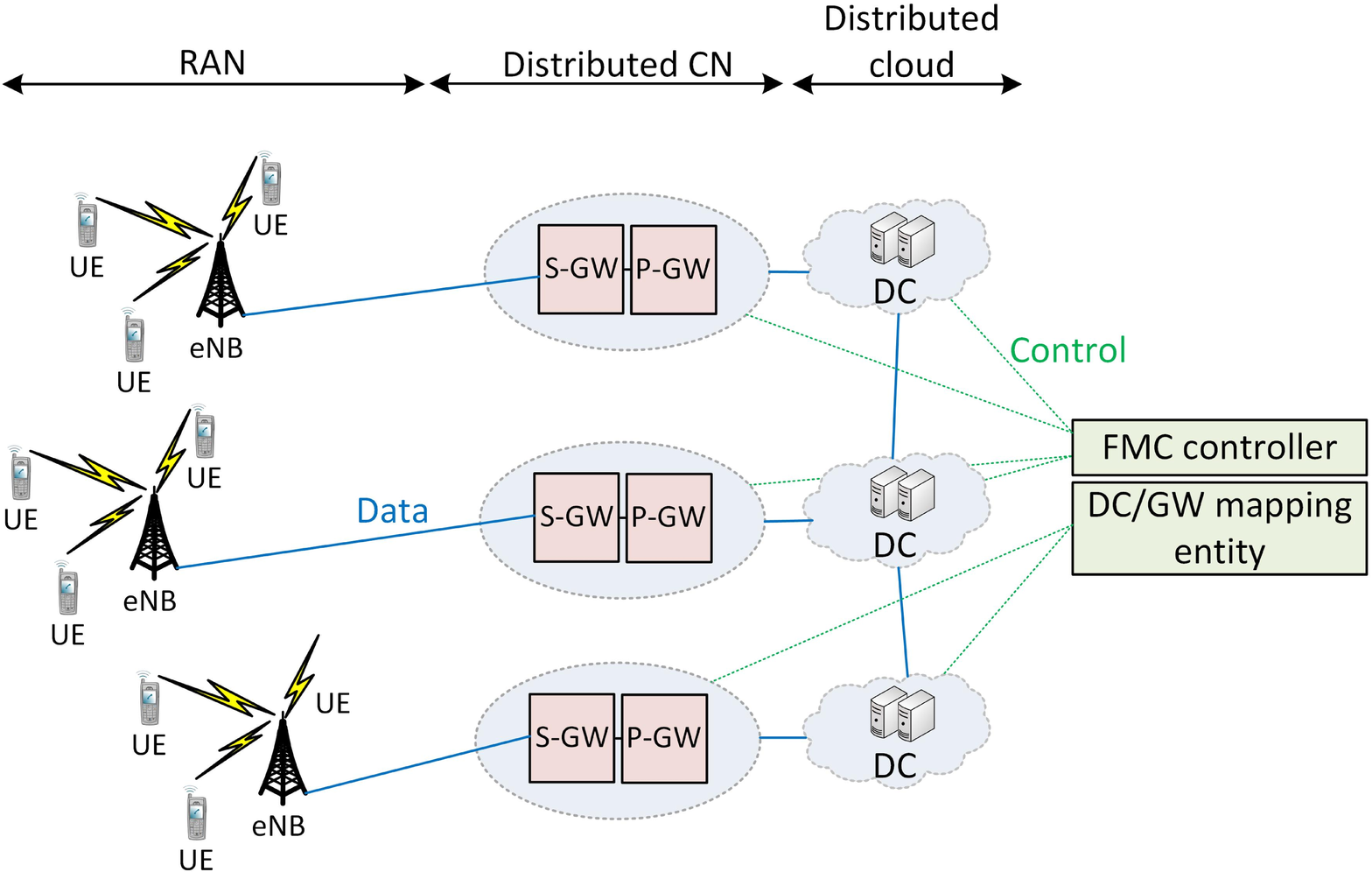} 
	\caption{The network architecture enabling FMC concept (centralized solution).}
	\label{fig:04}
\end{figure}

Analogously to the SCC, the MobiScud introduces a new control entity, a MobiScud control (MC), which interfaces with the mobile network, SDN switches and the cloud of the operator. Basically, the MC has two functionalities: 1) monitoring control plane signaling message exchange between mobile network elements to be aware of the UEs activity (e.g., handover) and 2) orchestrating and routing data traffic within SDN enabled transport network to facilitate the application offloading and the VM migration if the UE moves throughout the network.

\subsubsection{Follow me cloud (FMC)}
\label{sec214}
The key idea of the FMC is that the cloud services running at distributed data centers (DCs) follow the UEs as they roam throughout the network \cite{Taleb2013}\cite{Taleb2016} in the same way as in the case of the MobiScud. When compared to the previous MEC concepts, the computing/storage power is moved farther from the UEs; into the CN network of the operator. Nevertheless, while previous MEC concepts assume rather centralized CN deployment, the FMC leverages from the fact that the mobile operators need to decentralize their networks to cope with growing number of the UEs. In this respect, the centralized CN used in the current network deployment is assumed to be replaced by a distributed one as shown in Fig.~\ref{fig:04}. For a convenience of the mobile operators, the DC may be located at the same place as the distributed S/P-GWs. 

Similarly as the SCC and the MobiScud, the FMC introduces new entities into the network architecture; a DC/GW mapping entity and an FMC controller (FMCC). These can be either functional entities collocated with existing network nodes or a software run on any DC (i.e., exploiting NFV principles like the SCC or MobiScud concepts). The DC/GW mapping entity maps the DCs to the distributed S/P-GWs according to various metrics, such as, location or hop count between DC and distributed CN, in static or dynamic manner. The FMCC manages DCs' computation/storage resources, cloud services running on them, and decides which DC should be associated to the UE using the cloud services. The FMCC may be deployed either centrally (as shown in Fig.~\ref{fig:04}) or hierarchically \cite{Aissioui2015} with global FMCC (G-FMCC) and local FMCC (L-FMCC) for better scalability (controlled similarly  as in the SCC as explained in Section~\ref{sec211}). Note that the FMC itself may be also decentrally controlled by omitting the FMCC altogether. In such a case, the DCs coordinate themselves in a self-organizing manner.

\subsubsection{CONCERT}
\label{sec215}
\begin{figure}[t!]
	\centering
	\includegraphics[scale=0.09]{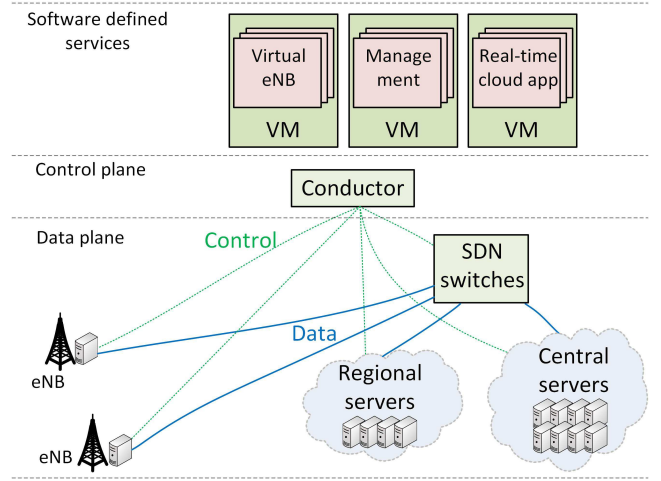} 
	\caption{CONCERT architecture.}
	\label{fig:05}
\end{figure}

A concept converging cloud and cellular systems, abbreviated as CONCERT, has been proposed in \cite{Liu2014}. The CONCERT assumes to exploit NFV principles and SDN technology like above-mentioned solutions. Hence, the computing/storage resources, utilized by both conventional mobile communication and cloud computing services, are presented as virtual resources. The control plain is basically consisted of a conductor, which is a control entity managing communication, computing, and storage resources of the CONCERT architecture. The conductor may be deployed centrally or in a hierarchical manner for better scalability as in the SCC or FMC. The data plain consists of radio interface equipments (RIEs) physically representing the eNB, SDN switches, and computing resources (see Fig.~\ref{fig:05}). The computing resources are used both for baseband processing (similarly as in C-RAN) and for handling an application level processing (e.g., for the application offloading). In all already described MEC concepts, the computation/storage resources have been fully distributed. The CONCERT proposes rather hierarchically placement of the resources within the network to flexibly and elastically manage the network and cloud services. In this respect, local servers with a low computation power are assumed to be located directly at the physical base station (e.g., similarly as in the SCC or the MMC) and, if the local resources are not sufficient, regional or even central servers are exploited as indicated in Fig.~\ref{fig:05}.

\subsection{ETSI MEC}
\label{sec22}
Besides all above-mentioned solutions, also ETSI is currently deeply involved in standardization activities in order to integrate the MEC into the mobile networks. In this regard, we briefly summarize the standardization efforts on the MEC within ETSI, describe reference architecture according to ETSI, and contemplate various options for the MEC deployment that are considered so far.

\subsubsection{Standardization of ETSI MEC}
\label{sec221}
Standardization of the MEC is still in its infancy, but drafts of specifications have already been released by ISG MEC. The terminology used in individual specifications relating to conceptual, architectural and functional elements is described in \cite{MEC001}. The main purpose of this document is to ensure the same terminology is used by all ETSI specifications related to the MEC. A framework exploited by ISG MEC for coordination and promotion of MEC is defined in proof of concept (PoC) specification \cite{MEC005}. The basic objectives of this document is to describe the PoC activity process in order to promote the MEC, illustrate key aspects of the MEC and build a confidence in viability of the MEC technology. Further, several service scenarios that should benefit from the MEC and proximity of the cloud services is presented in \cite{MEC004} \textcolor{black}{(see Section~\ref{Usecases} for more detail).} Moreover, technical requirements on the MEC to guarantee interoperability and to promote MEC deployment are introduced in \cite{MEC002}. The technical requirements are divided into generic requirements, service requirements, requirements on operation and management, and finally security, regulations and charging requirements.

\subsubsection{ETSI MEC reference architecture}
\label{sec222}
\begin{figure}[b!]
	\centering
	\includegraphics[scale=0.065]{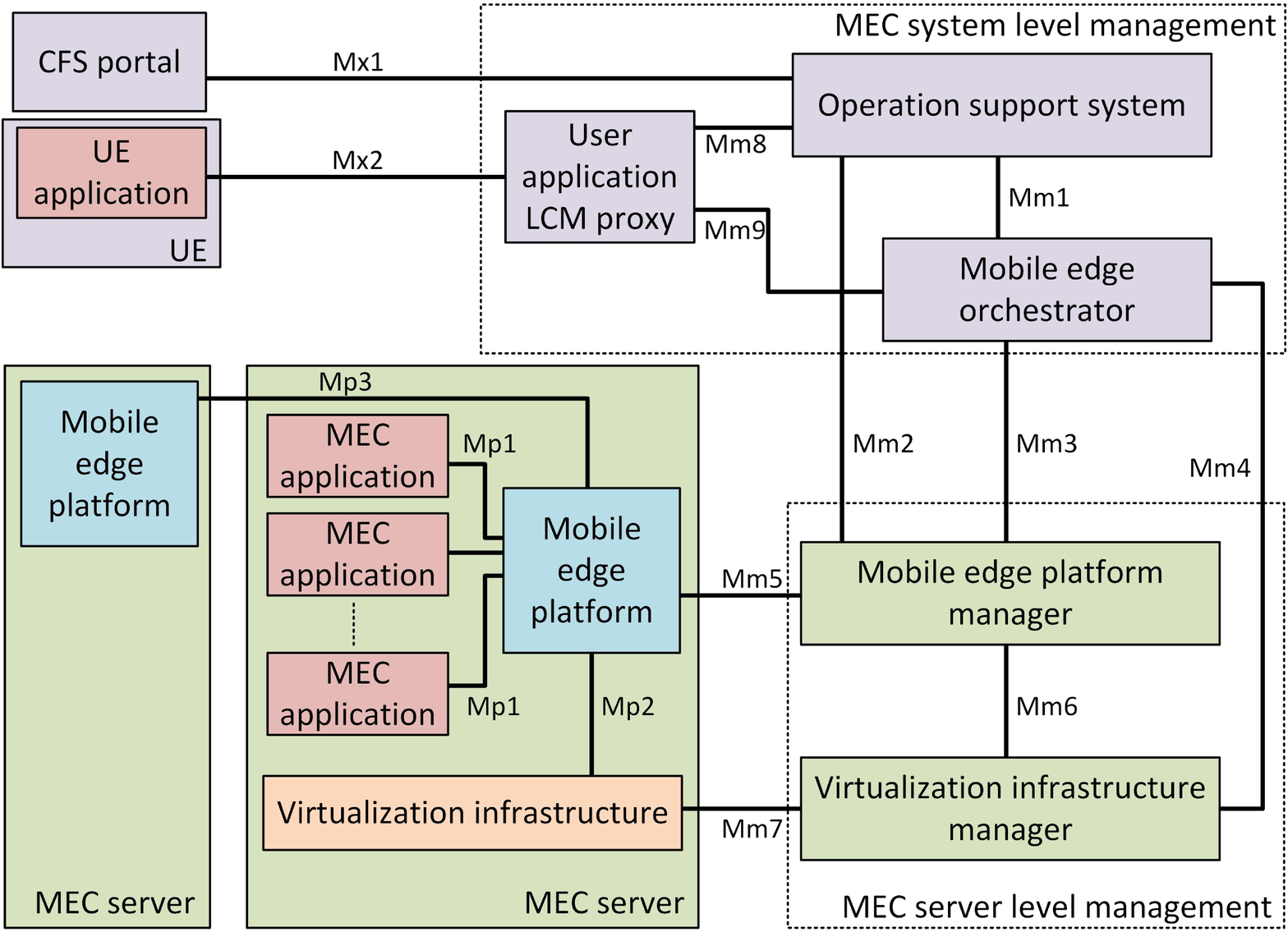} 
	\caption{MEC reference architecture \cite{MEC003}.}
	\label{fig:06}
\end{figure}

The reference architecture, described by ETSI in \cite{MEC003}, is composed of functional elements and reference points allowing interaction among them (see Fig.~\ref{fig:06}). Basically, the functional blocks may not necessarily represent physical nodes in the mobile network, but rather software entities running on the top of a virtualization infrastructure. The virtualization infrastructure is understood as a physical data center on which the VMs are run and the VMs represent individual functional elements. In this respect, it is assumed that some architectural features from ETSI NFV group, which runs in parallel to ETSI MEC, will be reused for the MEC reference architecture as well, since the basic idea of NFV is to virtualize all network node functions.

\begin{table*}[t]
\footnotesize
\caption{Comparison of existing MEC concepts.}
\label{tab:Tab2}
\centering
\renewcommand{\arraystretch}{1.2}
\begin{tabular}{|p{1.3cm}||p{1.5cm}|p{4cm}|p{4cm}|p{4cm}| }
\hline
		\bf MEC concept & \bf Control entity & \bf Control manner & \bf Control placement & \bf Computation/storage placement \\ \hline \hline
	 \emph{SCC} & SCM & Centralized, decentralized hierarchical (depending on SCM type and placement) & In RAN (e.g., at eNB) or in CN (e.g., SCM collocated with MME) & SCeNB, cluster of SCeNBs \\ \hline
	 \emph{MMC} & -  & Decentralized & MMC (eNB) & eNB \\ \hline
	 \emph{MobiScud} & MC & Decentralized & Between RAN and CN & Distributed cloud within RAN or close to RAN \\ \hline
	 \emph{FMC} & FMCC & Centralized, decentralized hierarchical (option with hierarchical FMCC), decentralized (option without FMC controller) & Collocated with existing node (e.g., node in CN) or run as software on DC & DC close or collocating with distributed CN \\ \hline
	 \emph{CONCERT} & Conductor & Centralized, decentralized hierarchical & N/A (it could be done in the same manner as in FMC concept) & eNB (RIE), regional and central servers \\ \hline
	 \emph{ETSI MEC} & Mobile edge orchestrator & Centralized & N/A (the most feasible option is to place control into CN & eNB, aggregation point, edge of CN \\ \hline
\end{tabular}
\end{table*}

As shown in Fig.~\ref{fig:06}, the MEC can be exploited either by a \emph{UE application} located directly in the UE, or by third party customers (such as commercial enterprise) via customer facing service (CFS) portal. Both the UE and the CFS portal interact with the MEC system through a \emph{MEC system level management}. The \emph{MEC system level management} includes a user application lifecycle management (LCM) proxy, which mediate the requests, such as initiation, termination or relocations of the UE's application within the MEC system to the operation support system (OSS) of the mobile operator. Then, the OSS decides if requests are granted or not. The granted requests are forwarded to a mobile edge orchestrator. The mobile edge orchestrator is the core functionality in the \emph{MEC system level management} as it maintains overall view on available computing/storage/network resources and the MEC services. \textcolor{black}{In this respect, the mobile edge orchestrator allocates the virtualized MEC resources to the applications that are about to be initiated depending on the applications requirements (e.g., latency). Furthermore, the orchestrator also flexibly scales down/up available resources to already running applications.}
 
The \emph{MEC system level management} is interconnected with a \emph{MEC server level management} constituting a mobile edge platform and a virtualization platform manager. The former one manages the life cycle of the applications, application rules and service authorization, traffic rules, etc. The latter one is responsible for allocation, management and release of the virtualized computation/storage resources provided by the virtualization infrastructure located within the MEC server. The MEC server is an integral part of the reference architecture as it represents the virtualized resources and hosts the MEC applications running as the VMs on top of the virtualization infrastructure.

\subsubsection{Deployment options of ETSI MEC}
\label{sec223}
As already mentioned in the previous subsection, the MEC services will be provided by the MEC servers, which have the computation and storage resources at their disposal. There are several options where the MEC servers can be deployed within the mobile network. The first option is to deploy the MEC server directly at the base station similarly as in case of the SCC or the MCC (see Section~\ref{sec211} and Section~\ref{sec212}). Note that in case of a legacy network deployment, such as 3G networks, the MEC servers may be deployed at 3G Radio Network Controllers as well \cite{MEC002}. The second option is to place the MEC servers at cell aggregation sites or at multi-RAT aggregation points that can be located either within an enterprise scenario (e.g., company) or a public coverage scenario (e.g., shopping mall, stadium, airport, etc.). The third option is to move the MEC server farther from the UEs and locate it at the edge of CN analogously to the FMC (Section~\ref{sec214}). 

Of course, selection of the MEC server deployment depends on many factors, such as, scalability, physical deployment constraints and/or performance criteria (e.g., delay). For example, the first option with fully distributed MEC servers deployment will result in very low latencies since the UEs are in proximity of the eNB and, hence, in proximity of the MEC server. Contrary, the UEs exploiting the MEC server located in the CN will inevitably experience longer latencies that could prevent a use of \textcolor{black}{real-time applications. An initial study determining where to optimally install the MEC servers within the mobile network with the primary objective to find a trade-off between installation costs and QoS measured in terms of latency is presented in \cite{Ceselli2015} and further elaborated in \cite{Ceselli2017}. Based on these studies, it is expected that, similarly as } in CONCERT framework (see Section~\ref{sec215}), the MEC servers with various computation power/storage capacities will be scattered throughout the network. Hence, the UEs requiring only a low computation power will be served by the local MEC servers collocated directly with the eNB, while highly demanding applications will be relegated to more powerful MEC servers farther from the UEs.

\subsection{Summary}
\label{sec23}
This section mutually compares the MEC concepts proposed in literature with the vision of the MEC developed under ETSI. There are two common trends followed by individual MEC solutions that bring cloud to the edge of mobile network. The first trend is based on virtualization techniques exploiting NFV’s principles. The network virtualization is a necessity in order to flexibly manage virtualized resources provided by the MEC. The second trend is a decoupling the control and data planes by taking advantage of SDN paradigm, which allows a dynamic adaptation of the network to changing traffic patterns and users requirements. \textcolor{black}{The use of SDN for the MEC is also in line with current trends in mobile networks \cite{Kreutz2015}-\cite{Jin2013}.} Regarding control/signaling, the MMC and MobiScud assume fully decentralize approach while the SCC, FMC, and CONCERT adopt either fully centralized control or hierarchical control for better scalability and flexibility. 

If we compare individual MEC concepts in terms of computation/storage resources deployment, the obvious effort is to fully distribute these resources within the network. Still, each MEC concept differs in the location, where the computation/storage resources are physically located. While the SCC, MMC and MobiScud assume to place the computation close to the UEs within RAN, the FMC solution considers integration of the DCs farther away, for example, in a distributed CN. On top of that, CONCERT distributes the computation/storage resources throughout the network in a hierarchical manner so that a low demanding computation application are handled locally and high demanding applications are relegated either to regional or central servers. Concerning ETSI MEC, there are also many options where to place MEC servers offering computation/storage resources to the UEs. The most probable course of action is that the MEC servers will be deployed everywhere in the network to guarantee high scalability of the computation/storage resources. The comparison of all existing MEC concepts is shown in Table~\ref{tab:Tab2}.

\section{\textcolor{black}{Introduction to computation offloading}}
\label{offloading}
\begin{figure}[b!]
	\centering
	\includegraphics[scale=0.105]{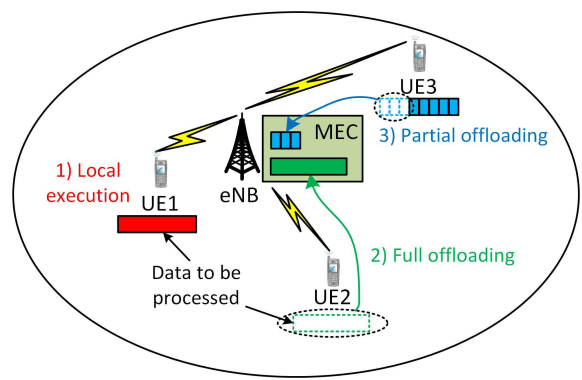} 
	\caption{Possible outcomes of computation offloading decision.}
	\label{fig:07}
\end{figure}

\textcolor{black}{From the user perspective, a critical use case regarding the MEC is a computation offloading as this can save energy and/or speed up the process of computation. In general, a crucial part regarding computation offloading is to decide whether to offload or not. In the former case, also a question is how much and what should be offloaded \cite{Zhang2012}. Basically, a decision on computation offloading may result in:}
\textcolor{black}{
\begin{itemize}[leftmargin=12pt]
\item
\textit{Local execution} - The whole computation is done locally at the UE (see Fig.~\ref{fig:07}). The offloading to the MEC is not performed, for example, due to unavailability of the MEC computation resources or if the offloading simply does not pay off.
\item	\textit{Full offloading} - The whole computation is offloaded and processed by the MEC.
\item	\textit{Partial offloading} - A part of the computation is processed locally while the rest is offloaded to the MEC. 
\end{itemize}}

\textcolor{black}{The computation offloading, and partial offloading in particular, is a very complex process affected by different factors, such as users’ preferences, radio and backhaul connection quality, UE capabilities, or cloud capabilities and availability \cite{Khan2014}. An important aspect in the computation offloading is also an application model/type since it determines whether full or partial offloading is applicable, what could be offloaded, and how. In this regard, we can classify the applications according to several criteria:}

\begin{itemize}[leftmargin=12pt]
\item
\textcolor{black}{\textit{Offloadability of application} - The application enabling code or data partitioning and parallelization (i.e., application that may be partially offloaded) can be categorized into two types. The first type of the applications is the app, which can be divided into $N$ offloadable parts that all can be offloaded (see Fig.~\ref{fig:11}a). Since each offloadable part may differ in the amount of data and required computation, it is necessary to decide which parts should be offloaded to the MEC. In the example given in Fig.~\ref{fig:11}a, 1\textsuperscript{st}, 2\textsuperscript{nd}, 3\textsuperscript{nd}, 6\textsuperscript{th}, and 9\textsuperscript{th} parts are processed locally while the rest is offloaded to the MEC. Notice that in the extreme case, this type of application may be fully offloaded to the MEC if no parts are processed by the UE. The second type of the applications is always composed of some non-offloadable part(s) that cannot be offloaded (e.g., user input, camera, or acquire position that needs to be executed at the UE \cite{Deng2016}) and $M$ offloadable parts. In Fig.~\ref{fig:11}b, the UE processes the whole non-offloadable part together with 2\textsuperscript{nd}, 6\textsuperscript{th}, and 7\textsuperscript{th} parts while the rest of the application is offloaded to the MEC.}

\begin{figure}[t!]
	\centering
	\includegraphics[scale=0.12]{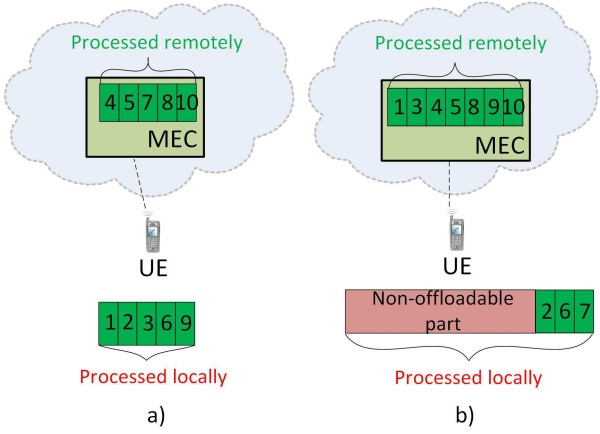} 
	\caption{An example of partial offloading for application without non-offloadable part(s) (a) and application with non-offloadable part (b).}
	\label{fig:11}
\end{figure}
\begin{figure}[b!]
	\centering
	\includegraphics[scale=0.11]{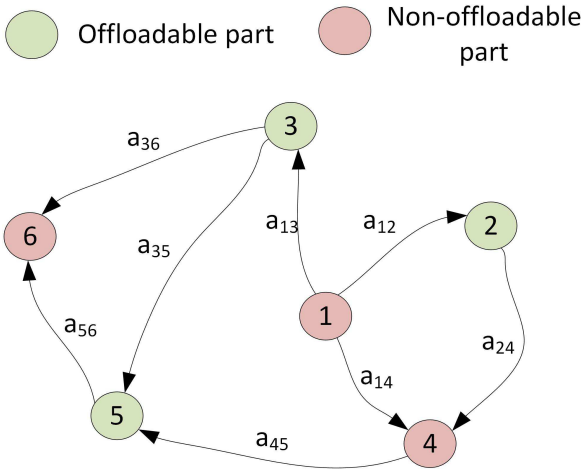} 
	\caption{Dependency of offloadable components \cite{Deng2016}.}
	\label{fig:12}
\end{figure}

\item
\textcolor{black}{\textit{Knowledge on the amount of data to be processed} - The applications can be classified according to the knowledge on the amount of data to be processed. For the first type of the applications (represented, e.g., by face detection, virus scan, etc.,) the amount of data to be processed is known beforehand. For the second type of the applications, it is not possible to estimate the amount of data to be processed as these are continuous-execution application and there is no way to predict how long they will be running (such as, online interactive games) \cite{Munoz2015}. It is obvious that decision on computation offloading could be quite tricky for continuous-execution application. }
\item
\textcolor{black}{\textit{Dependency of the offloadable parts} - The last criterion for classification of application to be offloaded is a mutual dependency of individual parts to be processed. The parts of the application can be either independent on each other or mutually dependent. In the former case, all parts can be offloaded simultaneously and processed in parallel. In the latter case, however, the application is composed of parts (components) that need input from some others and parallel offloading may not be applicable. Note that the relationship among individual components can be expressed by component dependency graph (CDG) or call graph (CG) (see, e.g., \cite{Zhang2015}\cite{Zhang2012}\cite{Deng2016}\cite{Mahmoodi2016}). The relationship among the components is illustrated in Fig.~\ref{fig:12}, where the whole application is divided into $M$ non-offloadable parts (1\textsuperscript{st}, 4\textsuperscript{th}, and 6\textsuperscript{th} part in Fig.~\ref{fig:12}) and $N$ offloadable parts (2\textsuperscript{nd}, 3\textsuperscript{rd}, and 5\textsuperscript{th} part in Fig.~\ref{fig:12}). In the given example, 2\textsuperscript{nd} and 3\textsuperscript{rd} part can be offloaded only after execution of the 1\textsuperscript{st} part while the 5\textsuperscript{th} part can be offloaded after execution of the 1\textsuperscript{st} - 4\textsuperscript{th} parts.} 
\end{itemize}

\textcolor{black}{The other important aspect regarding computation offloading is how to utilize and manage offloading process in practice. Basically, the UE needs to be composed of a code profiler, system profiler, and decision engine \cite{Flores2015}. The code profiler's responsibility is to determine what could be offloaded (depending on application type and code/data partitioned as explained above). Then, the system profiler is in charge of monitoring various parameters, such as available bandwidth, data size to be offloaded or energy spent by execution of the applications locally. Finally, decision engine determines whether to offload or not.}

\textcolor{black}{The next sections survey current research works focusing on following pivotal research topics: 1) decision on the computation offloading to the MEC, 2) efficient allocation of the computation resources within the MEC, and 3) mobility management for the moving users exploiting MEC services. Note that from now on we use explicitly the terminology according to ETSI standardization activities. Consequently, we use term MEC server as a node providing computing/storage resources to the UEs instead of DC, MMC, etc.}  

\section{Decision on computation offloading to MEC}
\label{sec3}
This section surveys current research related to the decision on the computation offloading to the MEC. The papers are divided into those considering either only the full offloading (Section~\ref{sec31}) or those taking into account also possibility of the partial offloading (Section~\ref{sec32}). 

\subsection{Full offloading}
\label{sec31}
The main objective of the works focused on the full offloading decision is to minimize an execution delay (Section~\ref{sec311}), to minimize energy consumption at the UE while predefined delay constraint is satisfied (Section~\ref{sec312}), or to find a proper trade-off between both the energy consumption and the execution delay (Section~\ref{sec313}). 

\subsubsection{Minimization of execution delay}
\label{sec311}
\begin{figure}[t!]
	%\centering
	\includegraphics[scale=0.096]{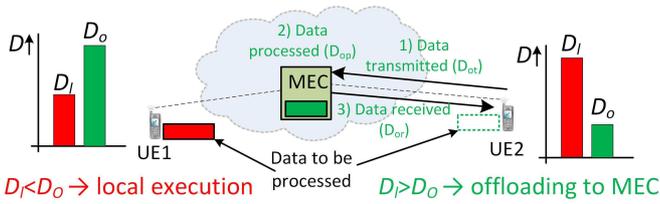} 
	\caption{The example of offloading decision aiming minimization of execution delay.}
	\label{fig:08}
\end{figure}

One of the advantages introduced by the computation offloading to the MEC is a possibility to reduce the execution delay ($D$). In case the UE performs all computation by itself (i.e., no offloading is performed), the execution delay ($D_l$) represents solely the time spent by the local execution at the UE. In case of the computation offloading to the MEC, the execution delay ($D_o$) incorporates three following parts: 1) transmission duration of the offloaded data to the MEC ($D_{ot}$), 2) computation/processing time at the MEC ($D_{op}$), and 3) time spent by reception of the processed data from the MEC ($D_{or}$). The simple example of the computation offloading decision based solely on the execution delay is shown in Fig.~\ref{fig:08}. It could be observed that the UE1 performs all computation locally since the local execution delay is significantly lower than expected execution delay for the computation offloading to the MEC (i.e., $D_l < D_o$). Contrary, a better alternative for the UE2 is to fully offload data to the MEC as the local execution would result in notable higher execution delay (i.e., $D_l > D_o$).

\begin{figure}[b!]
	\centering
	\includegraphics[scale=0.085]{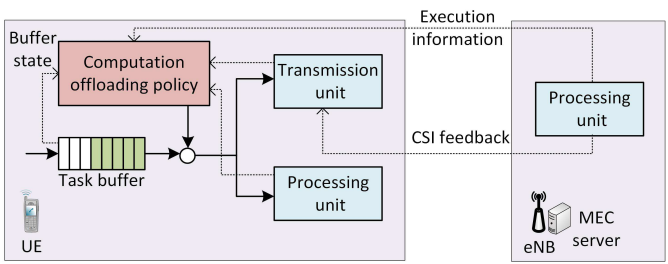} 
	\caption{Computation offloading considered in \cite{Liu2016} (CSI stands for channel state information).}
	\label{fig:09}
\end{figure}

\begin{figure*}[t!]
	\centering
	\includegraphics[scale=0.1]{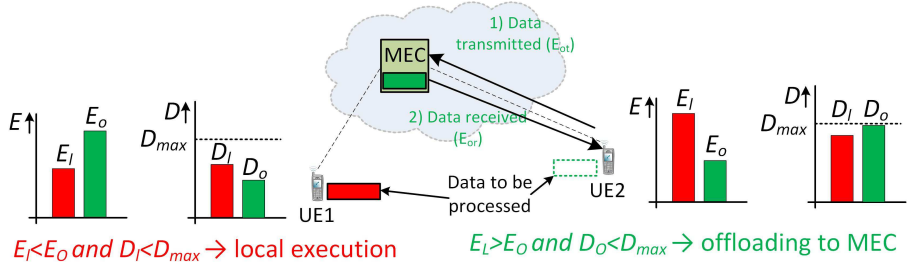} 
	\caption{The example of computation offloading decision based on energy consumption while satisfying execution delay constraint.}
	\label{fig:10}
\end{figure*}

The goal to minimize execution delay is pursued by the authors in \cite{Liu2016}. This is accomplished by one-dimensional search algorithm, which finds an optimal offloading decision policy according to the application buffer queuing state, available processing powers at the UE and at the MEC server, and characteristic of the channel between the UE and the MEC server. The computation offloading decision itself is done at the UE by means of a computation offloading policy module (see Fig.~\ref{fig:09}). This module decides, during each time slot, whether the application waiting in a buffer should be processed locally or at the MEC while minimizing the execution delay. The performance of the proposed algorithm is compared to the local execution policy (computation done always locally), cloud execution policy (computation performed always by the MEC server), and greedy offloading policy (UE schedules data waiting in the buffer whenever the local CPU or the transmission unit is idle). The simulation results show that the proposed optimal policy is able to reduce execution delay by up to 80\% (compared to local execution policy) and roughly up to 44\% (compared to cloud execution policy) as it is able to cope with high density of applications' arrival. The drawback of the proposed method is that the UE requires feedback from the MEC server in order to make the offloading decision, but the generated signaling overhead is not discussed in the paper.

Another idea aiming at minimization of the execution delay is introduced in \cite{Mao2016}. When compared to the previous study, the authors in \cite{Mao2016} also reduce application failure for the offloaded applications. The paper considers the UE applies dynamic voltage and frequency scaling (DVS) \cite{Zhang2013} and energy harvesting techniques \cite{Ulukus2015} to minimize the energy consumption during the local execution and a power control optimizing data transmission for the computation offloading. In this respect, the authors propose a low-complexity Lyapunov optimization-based dynamic computation offloading (LODCO) algorithm. The LODCO makes offloading decision in each time slot and subsequently allocates CPU cycles for the UE (if the local execution is performed) or allocates transmission power (if the computation offloading is performed). The proposed LODCO is able to reduce execution time by up to 64\% by offloading to the MEC. Furthermore, the proposal is able to completely prevent a situation when offloaded application would be dropped.

The drawback of both above-mentioned papers is that the offloading decision does not take into account energy consumption at the side of UE as fast battery depletion impose significant obstacle in contemporary networks. In \cite{Mao2016}, the energy aspect of the UE is omitted in the decision process since the paper assumes that the UEs exploit energy harvesting techniques. The harvesting technique, however, is not able to fully address energy consumption problem by itself.

\subsubsection{Minimization of energy consumption while satisfying execution delay constraint}
\label{sec312}
The main objective of the papers surveyed in this section is to minimize the energy consumption at the UE while the execution delay constraint of the application is satisfied. On one hand, the computation offloaded to the MEC saves battery power of the UE since the computation does not have to be done locally. On the other hand, the UE spends certain amount of energy in order to: 1) transmit offloaded data for computation to the MEC ($E_{ot}$) and 2) receive results of the computation from the MEC ($E_{or}$). The simple example of the computation offloading decision primarily based on the energy consumption is shown in Fig.~\ref{fig:10}. In the given example, the UE1 decides to perform the computation locally since the energy spent by the local execution ($E_l$) is significantly lower than the energy required for transmission/reception of the offloaded data ($E_0$). Contrary, the UE2 offloads data to the MEC as the energy required by the computation offloading is significantly lower than the energy spent by the local computation. Although the overall execution delay would be lower if the UE1 offloads computation to the MEC and also if the UE2 performs the local execution, the delay is still below maximum allowed execution delay constraint (i.e., $D_l < D_{max}$). Note that if only the execution delay would be considered for the offloading decision (as considered in Section~\ref{sec313}), both UEs would unnecessarily spent more energy.

The computation offloading decision minimizing the energy consumption at the UE while satisfying the execution delay of the application is proposed in \cite{Kamoun2015}. The optimization problem is formulated as a constrained Markov decision process (CMDP). To solve the optimization problem, two resource allocation strategies are introduced. The first strategy is based on an online learning, where the network adapts dynamically with respect to the application running at the UE. The second strategy is pre-calculated offline strategy, which takes advantage of a certain level of knowledge regarding the application (such as arrival rates measured in packets per slot, radio channel condition, etc.). The numerical experiments show that the pre-calculated offline strategy is able to outperform the online strategy by up to 50\% for low and medium arrival rates (loads). Since the offline resource allocation strategy proposed in \cite{Kamoun2015} shows its merit, the authors devise two addition dynamic offline strategies for the offloading \cite{Labidi2015}: deterministic offline strategy and randomized offline strategy. It is demonstrated that both offloading offline strategies can lead to significant energy savings comparing to the case when the computing is done solely at the UE (energy savings up to 78\%) or solely at the MEC (up to 15\%). 

A further extension of \cite{Labidi2015} from a single-UE to a multi-UEs scenario is considered in \cite{Labidi2015b}. The main objective is to jointly optimize scheduling and computation offloading strategy for each UE in order to guarantee QoE, fairness between the UEs, low energy consumption, and average queuing/delay constraints. The UEs that are not allowed to offload the computation make either local computation or stay idle. It is shown the offline strategy notably outperforms the online strategies in terms of the energy saving (by roughly 50\%). In addition, the energy consumed by individual UEs strongly depends on requirements of other UEs’ application.

Another offloading decision strategy for the multi-UEs case minimizing the energy consumption at the UEs while satisfying the maximum allowed execution delay is proposed in \cite{Barbarossa2013}. A decision on the computation offloading is done periodically in each time slot, during which all the UEs are divided into two groups. While the UEs in the first group are allowed to offload computation to the MEC, the UEs in the second group have to perform computation locally due to unavailable computation resources at the MEC (note that in the paper, the computation is done at the serving SCeNB). The UEs are sorted to the groups according to the length of queue, that is, according to the amount of data they need to process. After the UEs are admitted to offload the computation, joint allocation of the communication and computation resources is performed by finding optimal transmission power of the UEs and allocation of the SCeNB's computing resources to all individual UEs. The performance of the proposal is evaluated in terms of an average queue length depending on intensity of data arrival and a number of antennas used at the UEs and the SCeNB. It is shown that the more antennas is used, the less transmission power at the UEs is needed while still ensuring the delay constraint of the offloaded computation.
 
The main weak point of \cite{Barbarossa2013} is that it assumes only a single SCeNB and, consequently, there is no interference among the UEs connected to various SCeNBs. Hence, the work in \cite{Barbarossa2013} is extended in \cite{Sardellitti2014} to the multi-cell scenario with $N$ SCeNBs to reflect the real network deployment. Since the formulated optimization problem in \cite{Barbarossa2013} is no longer convex, the authors propose a distributed iterative algorithm exploiting Successive Convex Approximation (SCA) converging to a local optimal solution. The numerical results demonstrate that the proposed joint optimization of radio and computational resources significantly outperforms methods optimizing radio and computation separately. Moreover, it is shown that the applications with fewer amount of data to be offloaded and, at the same time, requiring high number of CPU cycles for processing are more suitable for the computation offloading. The reason is that the energy spent by the transmission/reception of the offloaded data to the MEC is significantly lower than the energy savings at the UE due to the computation offloading. The work in \cite{Sardellitti2014} is further extended in \cite{Sardellitti2014b} by a consideration of multi-clouds that are associated to individual SCeNBs. The results show that with an increasing number of the SCeNBs (i.e., with increasing number of clouds), the energy consumption of the UE proportionally decreases. 

The same goal as in previous paper is achieved in \cite{Zhang2016} by means of an energy-efficient computation offloading (EECO) algorithm. The EECO is divided into three stages. In the first stage, the UEs are classified according to their time and energy cost features of the computation to: 1) the UEs that should offload the computation to the MEC as the UEs cannot satisfy the execution latency constraint, 2) the UEs that should compute locally as they are able to process it by itself while the energy consumption is below a predefined threshold, and 3) the UEs that may offload the computation or not. In the second stage, the offloading priority is given to the UEs from the first and the third set determined by their communication channels and the computation requirements. In the third stage, the eNBs/SCeNBs allocates radio resources to the UEs with respect to given priorities. The computational complexity of the EECO is $O(max(I2+N, IK+N))$, where $I$ is the number of iterations, $N$ stands for amount of UEs, and $K$ represents the number of available channels. According to presented numerical results, the EECO is able to decrease the energy consumption by up to 15\% when compared to the computation without offloading. Further, it is proofed that with increasing computational capabilities of the MEC, the number of UEs deciding to offload the computation increases as well.

\subsubsection{Trade-off between energy consumption and execution delay}
\label{sec313}
The computation offloading decision for the multi-user multi-channel environment considering a trade-off between the energy consumption at the UE and the execution delay is proposed in \cite{Chen2016}. Whether the offloading decision prefers to minimize energy consumption or execution delay is determined by a weighing parameter. The main objective of the paper is twofold; 1) choose if the UEs should perform the offloading to the MEC or not depending on the weighing parameter and 2) in case of the computation offloading, select the most appropriate wireless channel to be used for data transmission. To this end, the authors present an optimal centralized solution that is, however, NP-hard in the multi-user multi-channel environment. Consequently, the authors also propose a distributed computation offloading algorithm achieving Nash equilibrium. Both the optimal centralized solution and the distributed algorithm are compared in terms of two performance metrics; 1) the amount of the UEs for which the computation offloading to the MEC is beneficial and 2) the computation overhead expressed by a weighing of the energy consumption and the execution delay. The distributed algorithm performs only slightly worse than the centralized one in both above-mentioned performance metrics. In addition, the distributed algorithm significantly outperforms the cases when all UEs compute all applications locally and when all UEs prefer computing at the MEC (roughly by up to 40\% for 50 UEs).

Other algorithm for the computation offloading decision weighing the energy consumption at the UE and the execution delay is proposed in \cite{Chen2015}. The main difference with respect to \cite{Chen2016} is that the authors in \cite{Chen2015} assume the computation can be offloaded also to the remote centralized cloud (CC), if computation resources of the MEC are not sufficient. The computation offloading decision is done in a sequential manner. In the first step, the UE decides whether to offload the application(s) to the MEC or not. If the application is offloaded to the MEC, the MEC evaluates, in the second step, if it is able to satisfy the request or if the computation should be farther relayed to the CC. The problem is formulated as a non-convex quadratically constrained quadratic program (QCQP), which is, however, NP-hard. Hence, a heuristic algorithm based on a semi-definite relaxation together with a novel randomization method is proposed. The proposed heuristic algorithm is able to significantly lower a total system cost (i.e., weighted sum of total energy consumption, execution delay and costs to offload and process all applications) when compared to the situation if the computation is done always solely at the UE (roughly up to 70\%) or always at the MEC/CC (approximately up to 58\%). 

The extension of \cite{Chen2015} from the single-UE to the multi-UEs scenario is presented in \cite{Chen2016b}. Since the multiple UEs are assumed to be connected to the same computing node (e.g., eNB), the offloading decision is done jointly with the allocation of computing and communication resources to all UEs. Analogously to \cite{Chen2015}, the proposal in \cite{Chen2016b} outperforms the case when computation is done always by the UE (system cost decreased by up to 45\%) and strategy if computation is always offloaded to the MEC/CC (system cost decreased by up to 50\%). Still, it would be useful to show the results for more realistic scenario with multiple computing eNBs, where interference among the UEs attached to different eNBs would play an important role in the offloading decision. Moreover, the overall complexity of the proposed solution is $O(N^6)$ per one iteration, which could be too high for a high number of UEs ($N$) connected to the eNB.

\subsection{Partial offloading}
\label{sec32}
\textcolor{black}{This subsection focuses on the works dealing with the partial offloading. We classify the research on works focused on minimization of the energy consumption at the UE while predefined delay constraint is satisfied (Section~\ref{sec321}) and works finding a proper trade-off between both the energy consumption and the execution delay (Section~\ref{sec322}).}

\subsubsection{Minimization of energy consumption while satisfying execution delay constraint}
\label{sec321}

This section focuses on the works aiming on minimization of the energy consumption while satisfying maximum allowable delay, similarly as in Section~\ref{sec312}. In \cite{Cao2015}, the authors consider the application divided into a non-offloadable part and $N$ offloadable parts as shown in Fig.~\ref{fig:11}b. The main objective of the paper is to decide, which offloadable parts should be offloaded to the MEC. The authors propose an optimal adaptive algorithm based on a combinatorial optimization method with complexity up to $O(2^N)$. To decrease the complexity of the optimal algorithm, also a sub-optimal algorithm is proposed reducing complexity to $O(N)$. The optimal algorithm is able to achieve up to 48\% energy savings while the sub-optimal one performs only slightly worse (up to 47\% energy savings). Moreover, it is shown that increasing SINR between the UE and the serving eNBs leads to more prominent energy savings.

The minimization of the energy consumption while satisfying the delay constrains of the whole application is also the main objective of \cite{Deng2016}. Contrary to \cite{Cao2015} the application in \cite{Deng2016} is supposed to be composed of several atomic parts dependable on each other, i.e., some parts may be processed only after execution of other parts \textcolor{black}{as shown in Fig.~\ref{fig:12} in Section~\ref{offloading}}. The authors formulate the offloading problem as $0-1$ programming model, where $0$ stands for the application offloading and $1$ represents the local computation at the UE. Nevertheless, the optimal solution is of a high complexity as there exists $2^N$ possible solutions to this problem (i.e., $O(2^NN^2)$). Hence, the heuristic algorithm exploiting Binary Particle Swarm Optimizer (BPSO) \cite{Kennedy1997} is proposed to reduce the complexity to $O(G.K.N^2)$, where $G$ is the number of iterations, and $K$ is the number of particles. The BPSO algorithm is able to achieve practically the same results as the high complex optimal solution in terms of the energy consumption. Moreover, the partial offloading results in more significant energy savings with respect to the full offloading (up to 25\% energy savings at the UE).

A drawback of both above papers focusing in detail on the partial computation offloading is the assumption of only single UE in the system. Hence, in \cite{Zhao2015}, the authors address the partial offloading decision problem for the multi-UEs scenario. With respect to \cite{Deng2016}\cite{Cao2015}, the application to be offloaded does not contain any non-offloadable parts and, in some extreme cases, the whole application may be offloaded if profitable \textcolor{black}{(i.e., the application is structured as illustrated in \ref{fig:11}a)}. The UEs are assumed to be able to determine whether to partition the application and how many parts should be offloaded to the MEC. The problem is formulated as a nonlinear constraint problem of a high complexity. As a consequence, it is simplified to the problem solvable by linear programming and resulting in the complexity $O(N)$ ($N$ is the number of UEs performing the offloading). If the optimal solution applying exhaustive search is used, 40\% energy savings are achieved when compared to the scenario with no offloading. In case of the heuristic low complex algorithm, 30\% savings are observed for the UEs. The disadvantage of the proposal is that it assumes the UEs in the system have the same channel quality and all of them are of the same computing capabilities. These assumptions, however, are not realistic for the real network.

A multi-UEs scenario is also assumed in \cite{You2016}, where the authors assume TDMA based system where time is divided into slots with duration of $T$ seconds. During each slot, the UEs may offload a part of their data to the MEC according to their channel quality, local computing energy consumption, and fairness among the UEs. In this regard, an optimal resource allocation policy is defined giving higher priority to those UEs that are not able to meet the application latency constraints if the computation would be done locally. After that, the optimal resource allocation policy with threshold based structure is proposed. In other words, the optimal policy makes a binary offloading decision for each UE. If the UE has a priority higher than a given threshold, the UE performs full computation offloading to the MEC. Contrary, if the UE has a lower priority than the threshold, it offloads only minimum amount of computation to satisfy the application latency constraints. Since the optimal joint allocation of communication and computation resources is of a high complexity, the authors also propose a sub-optimal allocation algorithm, which decouples communication and computation resource allocation. The simulation results indicate this simplification leads to negligibly higher total energy consumption of the UE when compared to the optimal allocation. The paper is further extended in \cite{You2016b}, where the authors show that OFDMA access enables roughly ten times higher energy savings achieved by the UEs comparing to TDMA system due to higher granularity of radio resources.

\textcolor{black}{In all above-mentioned papers on partial offloading, the minimization of UE's energy consumption depends on the quality of radio communication channel and transmission power of the UE. Contrary, in \cite{Wang2016c}, the minimization of energy consumption while satisfying execution delay of the application is accomplished through DVS technique. In this respect, the authors propose an energy-optimal partial offloading scheme that forces the UE adapt its computing power depending on maximal allowed latency of the application ($L_{MAX}$). In other words, the objective of the proposed scheme is to guarantee that the actual latency of the application is always equal to $L_{MAX}$. As a consequence, the energy consumption is minimized while perceived QoS by the users is not negatively affected.}    

\subsubsection{Trade-off between energy consumption and execution delay}
\label{sec322}

A trade-off analysis between the energy consumption and the execution delay for the partial offloading decision is delivered in \cite{Munoz2013}. Similarly as in \cite{Zhao2015}, the application to be offloaded contains only offloadable parts and in extreme case, the full offloading may occur (as explained in Section~\ref{sec32}). The offloading decision considers the following parameters: 1) total number of bits to be processed, 2) computational capabilities of the UE and the MEC, 3) channel state between the UE and the serving SCeNB that provides access to the MEC, and 4) energy consumption of the UE. The computation offloading decision is formulated as a joint optimization of communication and computation resources allocation. The simulation results indicate that the energy consumption at the UE decreases with increasing total execution time. This decrease, however, is notable only for small execution time duration. For a larger execution time, the gain in the energy savings is inconsequential. Moreover, the authors show the offloading is not profitable if the communication channel is of a low quality since a high amount of energy is spent to offload the application. In such situation, the whole application is preferred to be processed locally at the UE. With an intermediate channel quality, a part of the computation is offloaded to the MEC as this results in energy savings. Finally, if the channel is of a high quality, the full offloading is preferred since the energy consumption for data transmission is low while the savings accomplished by the computation offloading are high.

 \begin{table*}[t!]
\footnotesize
\caption{The comparison of individual papers addressing computation offloading decisions.}
\label{tab:Tab3}
\centering
\renewcommand{\arraystretch}{1}
\begin{tabular}{|p{0.32cm}|p{1.1cm}|p{2.5cm}|p{4cm}|p{1.3cm}|p{1.5cm}|p{2.4cm}|p{1.59cm}| }
\hline
	 &\bf Offloading type &\bf Objective &\bf Proposed solution &\bf No. of UE offloading & \bf Evaluation method &\bf Reduction of $D$/$E_{UE}$ wrt local computing &\bf Complexity of proposed algorithm \\ \hline
	\cite{Liu2016} & Full & 1) Minimize $D$
& One-dimensional search algorithm finding the optimal offloading policy & Single UE  & Simulations  & Up to \textbf{ 80\%} reduction of $D$  & N/A \\ \hline
	\cite{Mao2016} & Full &  1) Minimize $D$, 2) Minimize application failure  & Lyapunov optimization-based dynamic computation offloading & Single UE  & Theoretical verifications, simulations & Up to \textbf{64\%} reduction of $D$ & N/A \\ \hline
	\cite{Kamoun2015} & Full & 1) Minimize $E_{UE}$, 2) Satisfy $D$ constraint& Online learning allocation strategy, offline pre-calculated strategy  & Single UE  & Simulations & Up to \textbf{78\%} reduction of $E_{UE}$ & N/A \\ \hline
	\cite{Labidi2015} & Full & 1) Minimize $E_{UE}$, 2) Satisfy $D$ constraint & Deterministic and random offline strategies & Single UE &  Simulations & Up to \textbf{78\%} reduction of $E_{UE}$ & N/A \\ \hline
	\cite{Labidi2015b} & Full & 1) Minimize $E_{UE}$, 2) Satisfy $D$ constraint & Deterministic offline strategy, deterministic online strategy based on post-decision learning framework & Multi UEs &  Simulations & N/A & N/A \\ \hline
	 \cite{Barbarossa2013} & Full & 1) Minimize $E_{UE}$, 2) Satisfy $D$ constraint & Joint allocation of communication and computation resources & Multi UEs & Simulations & N/A & N/A \\ \hline
	\cite{Sardellitti2014} & Full & 1) Minimize $E_{UE}$, 2) Satisfy $D$ constraint & Distributed iterative algorithm exploiting Successive Convex Approximation (SCA) & Multi UEs & Simulations & N/A & N/A \\ \hline
	\cite{Sardellitti2014b} & Full & 1) Minimize $E_{UE}$, 2) Satisfy $D$ constraint & Distributed iterative algorithm exploiting Successive Convex Approximation (SCA) & Multi UEs & Simulations & N/A & N/A \\ \hline
	\cite{Zhang2016} & Full & 1) Minimize $E_{UE}$, 2) Satisfy $D$ constraint& Energy-efficient computation offloading (EECO) algorithm & Multi UEs  & Simulations & Up to \textbf{15\%} reduction of $E_{UE}$ & $O(max(I2+N, IK+N))$ \\ \hline
	 \cite{Chen2016} & Full & 1) Trade-off between $E_{UE}$ and $D$ & Computation offloading game & Multi UEs  & Analytical evaluations, simulations & Up to \textbf{40\%} reduction of $E_{UE}$ & N/A \\ \hline
	\cite{Chen2015} & Full & 1) Trade-off between $E_{UE}$ and $D$ & Heuristic algorithm based on semidefinite relaxation and randomization mapping method & Single UE  & Simulations & Up to \textbf{70\%} reduction of total cost & N/A \\ \hline
	\cite{Chen2016b} & Full & 1) Trade-off between $E_{UE}$ and $D$ & Heuristic algorithm based on semidefinite relaxation and randomization mapping method & Multi UEs  & Simulations & Up to \textbf{45\%} reduction of total cost  & $O(N^6)$ per iteration \\ \hline
	\cite{Cao2015} & Partial & 1) Minimize $E_{UE}$, 2) Satisfy $D$ constraint & Adaptive algorithm based on combinatorial optimization method & Single UE & Simulations & Up to \textbf{47\%} reduction of $E_{UE}$ & $O(N)$ \\ \hline
	 \cite{Deng2016} & Partial & 1) Minimize $E_{UE}$, 2) Satisfy $D$ constraint & Algorithm exploiting binary particle swarm optimizer & Single UE & Simulations & Up to \textbf{25\%} reduction of $E_{UE}$ & $O(G.K.N^2)$ \\ \hline
	\cite{Zhao2015} & Partial & 1) Minimize $E_{UE}$, 2) Satisfy $D$ constraint & Application and delay based resource allocation scheme & Multi UEs & Simulations & Up to \textbf{40\%}  reduction of $E_{UE}$ & $O(N)$ \\ \hline
	 \cite{You2016} & Partial & 1) Minimize $E_{UE}$, 2) Satisfy $D$ constraint & Optimal resource allocation policy with threshold based structure for TDMA system & Multi UEs & Simulations & N/A & N/A \\ \hline
	 \cite{You2016b} & Partial & 1) Minimize $E_{UE}$, 2) Satisfy $D$ constraint & Optimal resource allocation policy with threshold based structure for TDMA and OFDMA system & Multi UEs &  Simulations & N/A & $O(K+N)$ \\ \hline
\textcolor{black}{\cite{Wang2016c}} & \textcolor{black}{Partial} & \textcolor{black}{1) Minimize $E_{UE}$, 2) Satisfy $D$ constraint} & \textcolor{black}{Adapting computing power of the UE by means of DVS to achieve maximum allowed latency} & \textcolor{black}{Single UE} &  \textcolor{black}{Simulations} & \textcolor{black}{N/A} & \textcolor{black}{N/A} \\ \hline
	 \cite{Munoz2013} & Partial & 1) Trade-off between $E_{UE}$ and $D$ & Joint allocation of communication and computational resources & Single UE & Simulations & N/A & N/A \\ \hline
	\cite{Munoz2015} & Partial & 1) Trade-off between $E_{UE}$ and $D$ & Iterative algorithm finding the optimal value of the number of bits sent in uplink & Single UE & Analytical evaluations, simulations & Up to \textbf{97\%} reduction of $E_{UE}$ (SINR 45 dB, 4x4 MIMO) & N/A \\ \hline
	\cite{Munoz2014} & Partial & 1) Trade-off between $E_{UE}$ and $D$ & Joint allocation of communication and computational resources & Multi UEs & Simulations & Up to \textbf{90\%} reduction of $E_{UE}$  & N/A \\ \hline
	\cite{Mao2016b} & Partial & 1) Trade-off between $E_{UE}$ and $D$ & Lyapunov optimization-based dynamic computation offloading & Multi UEs & Simulations & Up  to \textbf{90\%} reduction of $E_{UE}$, up  to \textbf{98\%} reduction of $D$ & N/A \\ \hline
\end{tabular}
\end{table*} 

The study in \cite{Munoz2015} provides more in-depth theoretical analysis on trade-off between the energy consumption and the latency of the offloaded applications preliminarily handled in \cite{Munoz2013}. Moreover, the authors further demonstrate that a probability of the computation offloading is higher for good channel quality. With higher number of antennas (4x2 MIMO and 4x4 MIMO is assumed), the offloading is done more often and the energy savings at the UE are more significant when compared to SISO or MISO (up to 97\% reduction of energy consumption for 4x4 MIMO antenna configuration). Note that the same conclusion is also reached, e.g., in  \cite{Barbarossa2013}\cite{Sardellitti2014}. 

The main drawback in \cite{Munoz2013}\cite{Munoz2015} is that these papers consider only the single-UE scenario. A trade-of analysis between the energy consumption at the UE and the execution delay for the multi-UEs scenario is delivered in \cite{Munoz2014}. In case of the multi-UEs scenario, the whole joint optimization process proposed in \cite{Munoz2015} has to be further modified since both communication and computation resources provided by the MEC are shared among multiple UEs. In the paper, it is proven that with more UEs in the system, it takes more time to offload the application and it also lasts longer to process the application in the MEC. The reason for this phenomenon is quite obvious since less radio and computational resources remains for each UE. Still, up to 90\% of energy savings may be accomplished in multi-UE scenario.
 
A trade-off between the power consumption and the execution delay for the multi-UEs scenario is also tackled in \cite{Mao2016b}. The authors formulate a power consumption minimization problem with application buffer stability constraints. In this regard, the online algorithm based on Lyapunov optimization is proposed to decide on optimal CPU’s frequency for those UEs performing the local execution and to allocate transmission power and bandwidth to the UEs offloading the application to the MEC. The proposed algorithm is able to control the power consumption and the execution delay depending on the selected priority. The paper also demonstrates that the use of the MEC for the computation offloading is able to bring up to roughly 90\% reduction in the power consumption while the execution delay is reduced approximately by 98\%.

\subsection{Summary of works focusing on computation offloading decision}
\label{sec33}
A comparison of individual computation offloading strategies is illustrated in Table~\ref{tab:Tab3}. The majority of computation offloading decision algorithms aims to minimize the energy consumption at the UE ($E_{UE}$) while satisfying the execution delay ($D$) acceptable by the offloaded application or to find/analyse a trade-off between these two metrics. The papers indicate up to 90\% energy savings achievable by the computation offloading to the MEC and execution delay may be reduced even up to 98\%. Besides, all the papers evaluate the proposed solutions mostly by means of simulation (only several studies perform analytical evaluations). 

\section{Allocation of computing resources}
\label{sec4}
\begin{figure}[t!]
	\centering
	\includegraphics[scale=0.115]{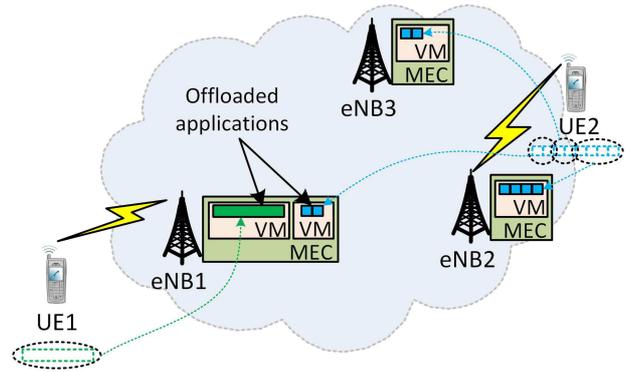} 
	\caption{An example of allocation of computing resources within the MEC.}
	\label{fig:13}
\end{figure}

If a decision on the full or partial offloading of an application to the MEC is taken (as discussed in previous section), a proper allocation of the computation resources has to be done. Similarly as in case of the computation offloading decision, the selection of computation placement is influenced by the ability of the offloaded application to be parallelized/partitioned. If the parallelization/partitioning of the application is not possible, only one physical node may be allocated for the computing since the application cannot be split into several parts (in Fig.~\ref{fig:13}, the UE1 offloads whole application to the eNB as this application cannot be partitioned). In the opposite case, the offloaded application may be processed by resources distributed over several computing nodes (in Fig.~\ref{fig:13}, the application offloaded by the UE2 is partitioned and processed by all three eNBs).

This section surveys the papers addressing the problem of a proper allocation of the computing resources for the applications that are going to be offloaded to the MEC (or in some cases to the CC, if the MEC computing resources are not sufficient). We categorize the research in this area into papers focusing on allocation of the computation resources at 1) a single computing node (Section~\ref{sec41}) and 2) multiple computing nodes (Section~\ref{sec42}).

\subsection{Allocation of computation resources at a single node}
\label{sec41}

\begin{figure}[b!]
	\centering
	\includegraphics[scale=0.083]{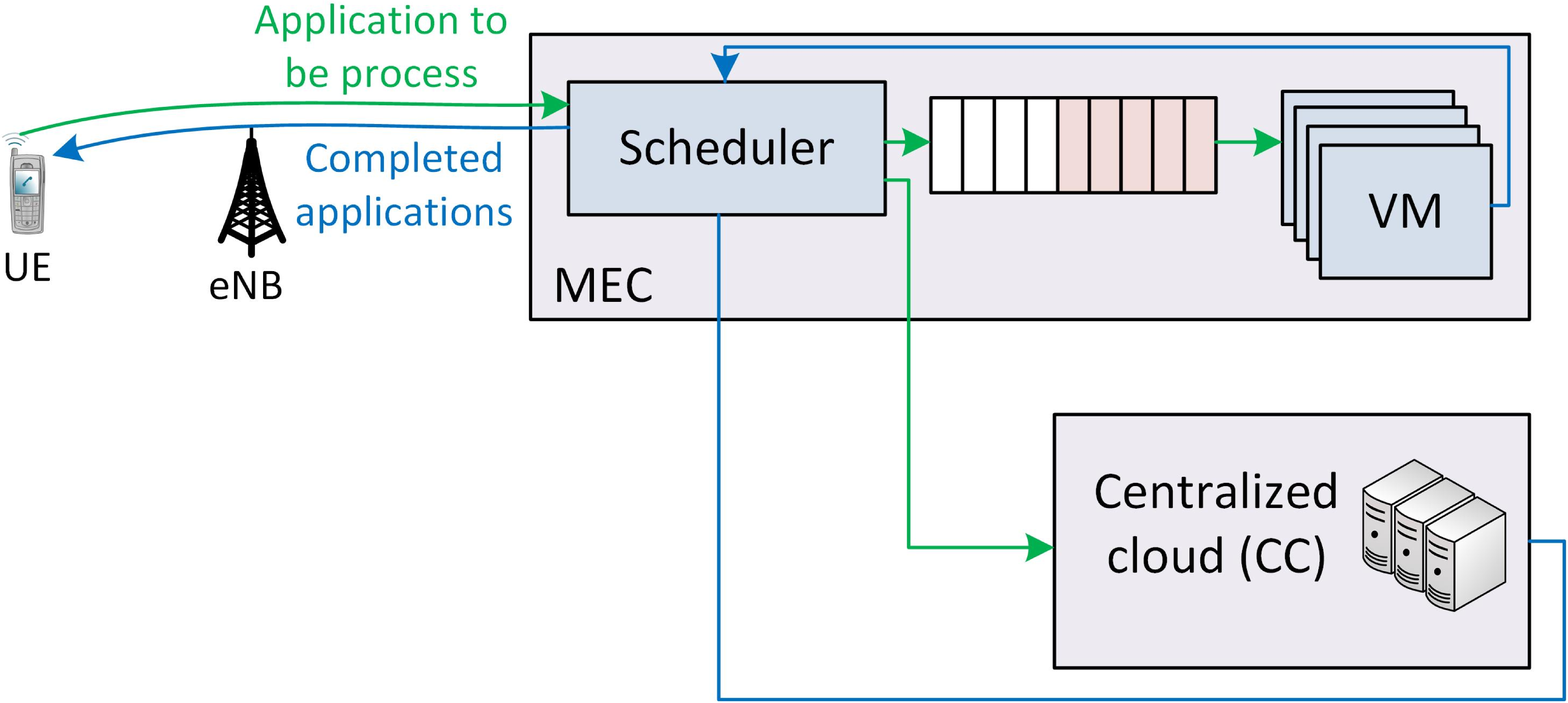} 
	\caption{Allocation of computation resources according to \cite{Zhao2015b}.}
	\label{fig:14}
\end{figure}

The maximization of the amount of the applications served by the MEC while satisfying the delay requirements of the offloaded applications is the main objective in \cite{Zhao2015b}. The decision where the individual applications should be placed depends on the applications priorities (derived from the application's delay requirements, i.e., the application with a low delay requirements has higher priority) and availability of the computing resources at the MEC. The basic principle for the allocation of computation resources is depicted in Fig.~\ref{fig:14}. The offloaded applications are firstly delivered to the local scheduler within the MEC. The scheduler checks if there is a computing node with sufficient computation resources. If there is a computing node with enough available resources, the VM is allocated at the node. Then the application is processed at this MEC node, and finally sent back to the UE (see Fig.~\ref{fig:14}). However, if the computation power provided by the MEC server is not sufficient, the scheduler delegates the application to the distant CC. In order to maximize the amount of applications processed in the MEC while satisfying their delay requirements, the authors propose a priority based cooperation policy, which defines several buffer thresholds for each priority level. Hence, if the buffer is full, the applications are sent to the CC. The optimal size of the buffer thresholds is found by means of low-complexity recursive algorithm. The proposed cooperation policy is able to increase the probability of the application completion within the tolerated delay by 25\%.

When compared to the previous paper, the general objective of \cite{Guo2016} is to minimize not only the execution delay but also the power consumption at the MEC. The paper considers a hot spot area densely populated by the UEs, which are able to access several MEC servers through nearby eNBs. To that end, an optimal policy is proposed using equivalent discrete MDP framework. However, this method results in a high communication overhead and high computational complexity with increasing number of the MEC servers. Hence, this problem is overcome by developing an application assignment index policy. In this respect, each eNB calculates its own index policy according to the state of its computing resources. Then, this index policy is broadcasted by all eNBs and the UE is able to select the most suitable MEC server in order to minimize both execution delay and power consumption. According to the results, the index policy is in the worst case by 7\% more costly than optimal policy in terms of system cost (note that system cost represents weighted execution delay and power consumptions of the MEC).
 
\begin{figure}[b!]
	\centering
	\includegraphics[scale=0.12]{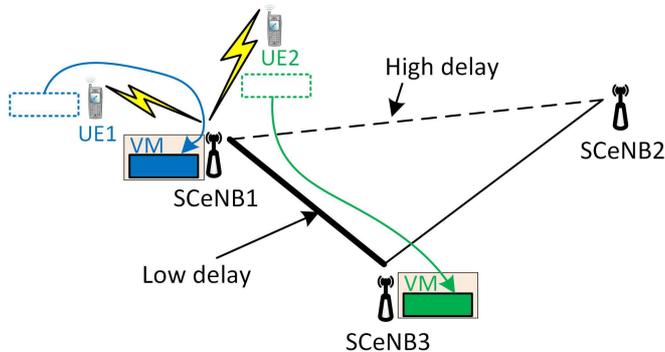} 
	\caption{An example of the VM allocation at single computing SCeNB according to \cite{Valerio2014}.}
	\label{fig:15}
\end{figure}

The minimization of the execution delay of the offloaded application is also the main goal in \cite{Valerio2014}. Nonetheless, with respect to \cite{Zhao2015b}\cite{Guo2016}, the other objectives are to minimize both communication and computing resource overloading and the VM migration cost (note that in \cite{Valerio2014}, the computing nodes are represented by the SCeNBs and the VM migration may be initiated due to the SCeNB’s shutdown). The whole problem is formulated as the VM allocation at the SCeNB and solved by means of MDP. An example of the VM allocation according to \cite{Valerio2014} is shown in Fig.~\ref{fig:15}, where the VM for the UE1 is allocated at the serving SCeNB1 while the UE2 has allocated the VM at the neighbouring SCeNB3. The SCeNB3 is preferred because of a high quality backhaul resulting in a low transmission delay of the offloaded data. The simulations show that with higher VM migration cost, the VM is preferred to be allocated at the serving SCeNB (i.e., the SCeNB closest to the UE) if this SCeNB has enough computation power. 

The main disadvantage of all above-mentioned approaches is that these do not consider more computing nodes within the MEC for single application in order to further decrease its execution delay.

\subsection{Allocation of computation resources at multiple nodes (federated clouds)}
\label{sec42}
When compared to the previous section, the allocation of computation resources at multiple computing nodes is considered here. The papers are split into subsections according to the main objective: 1) minimize execution delay and/or power consumption of computing nodes (Section~\ref{sec421}) and 2) balance both communication and computing loads (Section~\ref{sec421}).

\begin{figure}[t!]
	\centering
	\includegraphics[scale=0.105]{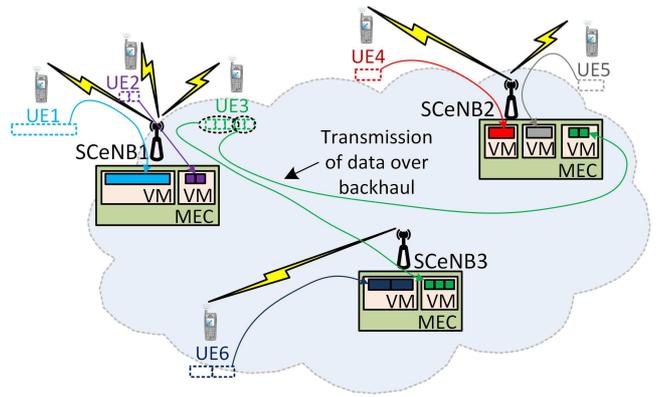} 
	\caption{An example of allocation of computation resources for individual UEs according to \cite{Tanzil2015}.}
	\label{fig:16}
\end{figure}

\subsubsection{Minimization of execution delay and/or power consumption of computing nodes}
\label{sec421}
The minimization of the execution delay by allocation of computing resources provided by the cluster of SCeNBs while avoiding to use the CC is proposed in \cite{Tanzil2015}. The cluster formation is done by means of a cooperative game approach, where monetary incentives are given to the SCeNBs if they perform the computation for the UEs attached to other SCeNBs. The coalition among the SCeNBs is formed for several time slots and then new coalitions may be created. The allocation of computation resources is done as shown in Fig.~\ref{fig:16}. Firstly, the serving SCeNB tries to serve their UEs on its own since this results in the shortest communication delay (e.g., in Fig.~\ref{fig:16} SCeNB1 allocates the computation resources to the UE1 and the UE2, etc.). Only if the SCeNB is not able to process the application on its own, it is forwarded to all SCeNBs in the same cluster (in Fig.~\ref{fig:16}, the computation for the UE3 is done at the SCeNB2 and the SCeNB3). The numerical results show that the proposed scheme is able to reduce the execution delay by up to 50\% when compared to the computation only at the serving SCeNB and by up to 25\% comparing to the scenario when all SCeNBs in the system participate in the computation. Unfortunately, the proposed approach does not address a problem of forming new coalitions and its impact on currently processed applications.

The selection of computing nodes can significantly influence not only the execution delay, as considered in \cite{Tanzil2015}, but also the power consumption of the computing nodes. Hence, the main objective of \cite{Oueis2014} is to analyze an impact of the cluster size (i.e., the amount of the SCeNBs performing computing) on both execution latency of the offloaded application and the power consumption of the SCeNBs. The analysis is done for different backhaul topologies (ring, tree, full mesh) and technologies (fiber, microwave, LTE). The authors demonstrate that full mesh topology combined with fiber or microwave connection is the most profitable in terms of execution latency (up to 90\% execution delay reduction). Contrary, a fiber backhaul in ring topology results in the lowest power consumption. Moreover, the paper shows that an increasing number of the computing SCeNBs does not always shorten execution delay. Quite the opposite, if a lot of SCeNBs process the offloading applications and the transmission delay becomes longer than the computing delay at the SCeNBs, the execution delay may be increased instead. Besides, with an increasing number of the computing SCeNBs, power consumption increases as well. Consequently, a proper cluster formation and the SCeNBs selection play a crucial part in system performance.   
 
The problem to find an optimal formation of the clusters of SCeNBs for computation taking into account both execution delay and power consumption of the computing nodes is addressed in \cite{Oueis2014b}. The paper proposes three different clustering strategies. The first clustering strategy selects the SCeNBs in order to minimize execution delay. Since all SCeNBs in the system model are assumed to be one hop away (i.e., full mesh topology is considered), basically all SCeNBs are included in the computation resulting in up to 22\% reduction of execution delay. This is due to the fact that the computation gain (and, thus, increase in the offloaded application processing) is far greater than the transmission delay. The objective of the second clustering strategy is to minimize overall power consumption of the cluster. In this case, only the serving SCeNB is preferred to compute, thus, any computation at the neighbouring SCeNBs is suppressed to minimize power consumption of the SCeNBs (up to 61\% reduction of power consumption is observed). This, however, increases overall latency and high variations of the computation load. The last clustering strategy aims to minimize the power consumption of each SCeNB in the cluster, since the power consumptions of the individual SCeNBs is highly imbalanced in the second strategy. 

While in \cite{Oueis2014b} the optimal clustering of the SCeNBs is done only for single UE, the multi UEs scenario is assumed in \cite{Oueis2015}. When compared to the previous paper, whenever the UE is about to offload data for the computation, the computing cluster is assigned to it. Consequently, each UE has assigned different cluster size depending on the application and the UE's requirements. The core idea of the proposal is to jointly compute clusters for all active users' requests simultaneously to being able efficiently distribute computation and communication resources among the UEs and to achieve higher QoE. The main objective is to minimize the power consumption of the clusters while guaranteeing required execution delay for each UE. The joint clusters optimization is able to significantly outperform the successive cluster optimization (allocation of the clusters are done subsequently for each UE), the static clustering (equal load distribution among SCeNBs) and no clustering (computation is done only by the serving SCeNB) in terms of the users' satisfaction ratio (up to 95\% of UEs is satisfied). On the other hand, the average power consumption is significantly higher when compared to "no clustering" and "successive clusters optimization" scenarios.

\begin{figure}[b!]
	\centering
	\includegraphics[scale=0.093]{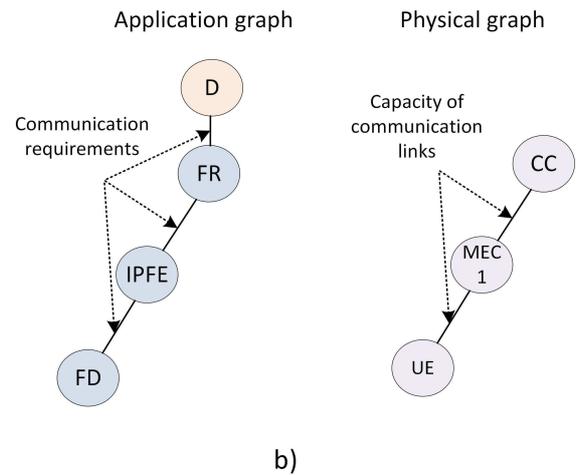} 
	\caption{An example of application and physical graph according to \cite{Wang2016} (FD - Face detection, IPFE - Image processing and feature extraction, FR - Face recognition, D - Database).}
	\label{fig:17}
\end{figure}

Similar as in \cite{Oueis2015}, the multi-UE cluster allocation is assumed in \cite{Oueis2015b}, but the cluster formation is done jointly with the UEs scheduling. The proposed resource allocation process is split into two steps similarly as proposed in \cite{Tanzil2015}. In the first step, labeled as \emph{local computational resource allocation}, each SCeNB allocates its computational resources to their own UEs according to specific scheduling rules, such as application latency constraint, computation load or minimum required computational capacity. In the second step, labelled as \emph{establishment of computing clusters}, the computation clusters are created for each UE that cannot be served by its serving SCeNB. The authors propose three algorithm realizations differing in applications prioritization (e.g., earliest deadline first or according to computation size of application) and the objective (minimization of power consumption or execution latency similarly as, e.g., in \cite{Oueis2015}). The simulations illustrate that there could be found the algorithm realization resulting in the users satisfaction ratio above 95\% while keeping a moderate power consumption of all computing nodes.

\begin{table*}[t!]
\footnotesize
\caption{The comparison of individual papers addressing allocation of computation resources for application/data already decided to be offloaded.}
\label{tab:Tab4}
\centering
\renewcommand{\arraystretch}{1}
\begin{tabular}{|p{0.42cm}|p{1.7cm}|p{3.2cm}|p{4cm}|p{2.1cm}|p{1.2cm}|p{2.4cm}|}
\hline
	 &\bf No. of computing nodes for each application & \bf Objective & \bf Proposed solution & \bf Computing nodes & \bf Evaluation method & \bf Results \\ \hline
	\cite{Zhao2015b} & Single node & 1) Maximize the amount of served applications, 2) Satisfy $D$ constraint & Priority based cooperation policy  & MEC servers (e.g., at the eNB or agg. point), CC & Simulations & \textbf{25\%} reduction of $D$ wrt offloading only to the MEC   \\ \hline
	 	\cite{Guo2016} & Single node & 1) Minimize $D$, 2) Minimize $E_C$ & Optimal online application assignment policy using equivalent discrete MDP framework & MEC servers (e.g., at the eNB or agg. point), CC & Simulations & N/A \\ \hline
	\cite{Valerio2014} & Single node & 1) Minimize $D$, 2) Minimize overloading of communication and computing resources, 3) Minimize VM migration cost & Optimal allocation policy obtained by solving MDP using linear programing reformulation & SCeNBs & Simulations & N/A \\ \hline
	 \cite{Tanzil2015} & Multiple nodes & 1) Minimize $D$, 2) Avoid to use the CC due to high delay & Formation of collaborative coalitions by giving monetary incentives to the SCeNBs  & SCeNBs & Simulations & Up to \textbf{50\%} reduction of $D$ wrt single computing SCeNB \\ \hline
	 \cite{Oueis2014} & Multiple nodes & 1) Analyze the impact of different network topologies and technologies on execution delay and power consumption & N/A & SCeNBs & Simulations & Up to 90\% reduction of $D$ wrt single computing SCeNB \\ \hline
	\cite{Oueis2014b} & Multiple nodes & 1) Minimize $D$, 2) Minimize $E_C$ & Three clustering strategies minimizing delay, power consumption of the cluster and power consumption of the SCs & SCeNBs & Simulations & \textbf{22\%} reduction of $D$, 61\% reduction of $E_C$ \\ \hline
	 \cite{Oueis2015} & Multiple nodes & 1) Minimize $D$, 2) Minimize $E_C$& Joint cluster formation for all active users’ requests simultaneously & SCeNBs & Simulations & Up to \textbf{95\%} of UE are satisfied (for max. 5 UEs) \\ \hline
	\cite{Oueis2015b} & Multiple nodes & 1) Minimize $D$, 2) Minimize $E_C$& Joint cluster formation for all active users’ requests simultaneously together with users scheduling & SCeNBs & Simulations & Up to \textbf{95\%} of UE are satisfied (for max. 5 UEs) \\ \hline
	 \cite{Vondra2014} & Multiple nodes & 1) Balance communication and computation load of computing nodes, 2) Satisfy execution delay requirement & ACA algorithm assuming jointly computation and communication loads  & SCeNBs & Simulations & \textbf{100\%} satisfaction ratio for up to 6 offloaded tasks/s  \\ \hline
	 \cite{Wang2016} & Multiple nodes & 1) Balance communication and computation load of computing nodes, 2) Minimize resource utilization& Online approximation algorithms with polynomial-logarithmic (poly-log) competitive ratio for tree application graph placement& UE, eNB, CC & Simulations & Reduction of resource utilization up to \textbf{10\%} \\ \hline
	 	\end{tabular}
\end{table*}

\subsubsection{Balancing of communication and computation load}
\label{sec422}
In the previous section, the allocation of computing resources is done solely with purpose to minimize the execution delay and/or the power consumption of the computing nodes. This could, however, result in unequal load distribution among individual computing nodes and backhaul overloading. The balancing of communication and computation load of the SCeNBs while satisfying the delay requirement of the offloaded application is addressed in \cite{Vondra2014}. To this end, an Application Considering Algorithm (ACA) selecting suitable SCeNBs according to the current computation and communication load of the SCeNBs is proposed. The ACA exploits knowledge of the offloaded application's requirements (i.e., the number of bytes to be transferred and the maximum latency acceptable by the application/user). The selection of the SCeNBs for the computation is done in a static way prior to the offloading to avoid expensive VMs migration. The performance evaluation is done for two backhauls, low throughput ADSL and high quality gigabit passive optical network (GPON). The proposed ACA algorithm is able to satisfy 100\% of the UEs as long as number of offloaded tasks per second is up to 6. Moreover, the paper shows that tasks parallelization helps to better balance computation load.  

The main objective to balance the load (both communication and computation) among physical computing nodes and, at the same time, to minimize the resource utilization of each physical computing node (i.e., reducing sum resource utilization) is also considered in \cite{Wang2016}. The overall problem is formulated as a placement of application graph onto a physical graph. The former represents the application where nodes in graph correspond to individual components of the application and edges to the communication requirements between them. The latter represents physical computing system, where the nodes in graph are individual computing devices and edges stands for the capacity of the communication links between them (see the example of application and physical graphs in Fig.~\ref{fig:17} for the face recognition application). The authors firstly propose the algorithm finding the optimal solution for the linear application graph and, then, more general online approximation algorithms. The numerical results demonstrate that the proposed algorithm is able to outperform two heuristic approaches in terms of resource utilization by roughly 10\%.

\subsection{Summary of works dealing with allocation of computing resources}
\label{sec43}
The comparison of individual methods addressing allocation of the computation resources within the MEC is shown in Table~\ref{tab:Tab4}. The main objective of the studies dealing with the allocation of computation resources is to minimize the execution delay of the offloaded application ($D$). In other words the aim is to ensure QoS to the UEs in order to fully exploit proximity of the MEC with respect to the computing in faraway CC. Moreover, several studies also focus on minimization of the energy consumption of computing nodes ($E_C$). In addition, some limited effort has been focused on balancing of computing and communication load to more easily satisfy the requirements on execution delay and/or to minimize overall resources utilization.

A common drawback of all proposed solutions is that only simulations are provided to demonstrate proposed solutions for allocation of MEC computing resources. Moreover, all papers disregard mobility of the UEs. Of course, if the UEs are fixed, individual proposal yield a satisfactory execution delay and/or power consumption at the computing nodes. Nevertheless, if the UE moves far away from the computing nodes, this could result in significant QoS degradation due to long transmission latency and extensive user’s dissatisfaction. This issue is addressed in the subsequent section targeting mobility management for the MEC.

\section{Mobility management for MEC}
\label{sec5}
In the conventional mobile cellular networks, a mobility of users is enabled by handover procedure when the UE changes the serving eNB/SCeNB as it roams throughout the network to guarantee the service continuity and QoS. Analogously, if the UE offloads computation to the MEC, it is important to ensure the service continuity. In fact, there are several options how to cope with the mobility of UEs. The first option, applicable only for the UEs with a low mobility (e.g. within a room), is to adapt transmission power of the eNB/SCeNB during the time when the offloaded application is processed by the MEC (Section~\ref{sec51}). If the UE performs handover to the new serving eNB/SCeNB despite of the power control, the service continuity may be guarantee either by the VM migration (i.e., the process during which the VM run at the current computing node(s) is migrated to another, more suitable, computing node(s) as discussed in Section~\ref{sec52}) or by selection of a new communication path between the UE and the computing node (Section~\ref{sec53}).

\subsection{Power control}
\label{sec51}
\begin{figure}[t!]
	\centering
	\includegraphics[scale=0.12]{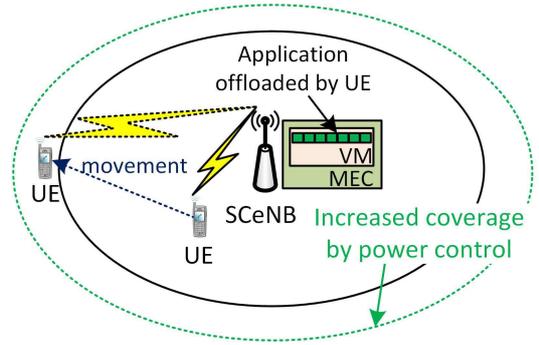} 
	\caption{Principle of CaPC according to \cite{Mach2014}\cite{Mach2016}.}
	\label{fig:18}
\end{figure}

In case when the UEs' mobility is low and limited, e.g., when the UEs are slowly moving inside a building, a proper setting of the transmission power of the serving and/or neighboring SCeNBs can help to guarantee QoS. This is considered in \cite{Mach2014}, where the authors propose a cloud-aware power control (CaPC) algorithm helping to manage the offloading of real-time applications with strict delay requirements. The main objective of the CaPC is to maximize the amount of the offloaded applications processed by the MEC with a given latency constrain. This is achieved by an adaptation of the transmission power of the SCeNBs so that the handover to a new SCeNB is avoided if possible (see the basic principle in Fig.\ref{fig:18} where the moving UE remains connected to the same SCeNB as its transmission power is increased). The CaPC is composed of coarse and fine settings of the SCeNBs’ transmission power. The purpose of the coarse setting is to find an optimal default transmission power $P_{t,def}$, which is applied if all of the UEs attached to the SCeNB are idle. Setting of the $P_{t,def}$ depends on the power level received by the serving SCeNB from the most interfering neighboring SCeNB and the interference generated by the eNBs. The fine setting consists in a short-term adaptation of the SCeNB's transmission power when the UE would not be able to receive the offloaded application from the cloud due to low SINR. If the CaPC is utilized, up to 95\% applications computed at the SCeNBSs are successfully delivered back to the UE with satisfying delay. Contrary, a conventional, non-cloud-aware, power control is able to successfully deliver only roughly 80\% of offloaded applications.
 
The main disadvantage of the CaPC presented in \cite{Mach2016} is that the time when the fine adjustment of the transmission power is triggered ($\Delta_t$) is the same for all SCeNBs and UEs independently on the channel quality (i.e., SINR). As a consequence, the CaPC may be triggered too late when sufficient SINR cannot be guaranteed in due time to successfully deliver the offloaded application back to the UE. This problem is addressed in \cite{Mach2016}, where the $\Delta_t$ is set individually for each UEs depending on its current channel quality. The proposed algorithm finds $\Delta_t$ by iterative process when $\Delta_t$ is adapted after each application is successfully delivered back to the UE. This way, the amount of successfully delivered applications is increased up to 98\%, as demonstrated by simulations.

\subsection{VM migration}
\label{sec52}
If the UEs’ mobility is not limited, as considered in Section~\ref{sec51}, and power control is no longer sufficient to keep the UE at the same serving eNB/SCeNB, a possibility to initiate the VM migration should be contemplated in order to guarantee the service continuity and QoS requirements. \textcolor{black}{On one hand, the VM migration has its cost ($Cost_M$) representing the time required for the VM migration and backhaul resources spent by transmission of the VM(s) between the computing nodes. On the other hand, there is a gain if the VM migration is initiated ($Gain_{M}$) since the UE can experience lower latency (data is processed in UE's vicinity) and backhaul resources do not have to be allocated for transmission of the computation results back to the UE.}

A preliminary analysis how the VM migration influences performance of the UE is tackled in \cite{Taleb2013b}. The authors describe analytical model based on Markov chains. Without the VM migration, a probability that the UE is connected to the optimal MEC decreases with increasing number of hops between the UE and the eNB, where the service is initially placed. This also results in increasing delay. Contrary, the connection of the UE to the optimal MEC server results in the lowest delay but at the high cost of the migration. The reason for this phenomenon is that the VM migration should be ideally initiated after each handover performed by the UE to keep minimum delay.

\begin{figure}[t!]
	\centering
	\includegraphics[scale=0.1]{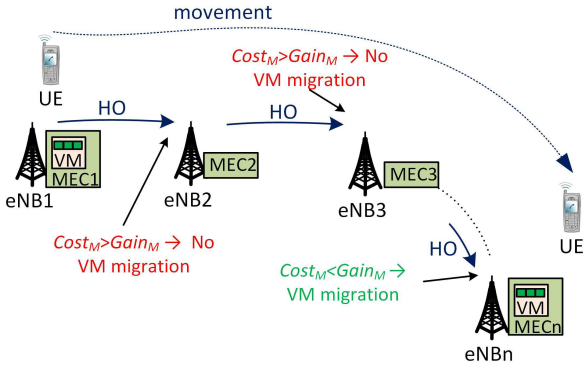} 
	\caption{VM migration principle according to \cite{Ksentini2014}.}
	\label{fig:19}
\end{figure}

While the previous paper is more general and focused on preliminary analysis regarding the VM migration, the main objective of \cite{Ksentini2014} is to design a proper decision policy determining whether to initiate the VM migration or not. As discussed above, there is a trade-off between the migration cost ($Cost_M$) and migration gain ($Gain_{M}$). The authors formulate the VM migration policy as a Continuous Time MDP (CTMDP) and they try to find an optimal threshold policy when the VM migration is initiated. Consequently, after each handover to the new eNB, the optimal threshold policy decides whether the VM migration should be initiated or not. An example of this principle is shown in Fig.~\ref{fig:19}, where the UE exploiting the MEC1 moves from the eNB1 to the eNBn. While the conventional radio handover is performed whenever the UE crosses cell boundaries, the VM migration is initiated after handover to the eNBn is performed since $Cost_M < Gain_{M}$. Simulations show, that the proposed optimal policy always achieves the maximum expected gain if compared to never migrate strategy (i.e., the computation is located still at the same MEC) and the scheme when the VM migration is performed after a specific number of handovers (10 handovers is set in the paper).

\textcolor{black}{A proper trade-off between VM migration cost ($Cost_M$) and VM migration gain ($Gain_M$) is also studied in \cite{Sun2016}. The paper proposes a Profit Maximization Avatar Placement  (PRIMAL) strategy deciding whether the VM should be migrated or not. Since the PRIMAL problem is NP-hard, the authors use Mixed-Integer Quadratic Programming tool to find the heuristic solution. The proposed solution is able to significantly reduce execution delay when compared to the situation with no migration (roughly by 90\%) while reducing the migration cost approximately by 40\%. When compared to \cite{Ksentini2014}, the authors also show the influence of $\alpha$ parameter weighing $Cost_M$ and $Gain_M$. Basically, with increasing $\alpha$, the migration cost is decreasing (i.e., migration is not done so frequently), but at the cost of higher execution delay.}    

An optimal threshold policy for the VM migration is also considered in \cite{Wang2014}. The problem is again formulated as the MDP and the VM migration is initiated always if the state of the UE is bounded by a particular set of thresholds. The state of the UE is defined as the number of hops between the eNB to which the UE is connected and the location of the MEC server where the computing service is running (in the paper labelled as the offset). The main objective of the paper is to minimize the overall sum cost by the optimal VM migration decision (i.e., the VM migration is performed if $Cost_M < Gain_{M}$  as explained earlier). The authors proof the existence of the optimal threshold policy and propose an iterative algorithm in order to find the optimal thresholds for the VM migration. The time complexity of the algorithm is $O(|M|N)$, where $M$ and $N$ is the maximum negative and positive offset, respectively. The performed results proof the optimal threshold policy is able to always outperform "never migrate" or "always migrate" strategies in terms of the sum cost. 

The main drawback of \cite{Ksentini2014}\cite{Wang2014} is that these assume simple 1D mobility model. More general setting for the VM migration is contemplated in \cite{Wang2015b}, where 2D mobility and real mobility traces are assumed. The authors formulate a sequential decision making problem for the VM migration using MDP and define algorithm for finding optimal policy with the complexity $O(N^3)$, where $N$ is the number of states (note that the state is defined as the number of hops between the eNB to which the UE is connected and the location of the MEC server analogously to \cite{Wang2014}). Since the proposed optimal VM migration strategy is too complex, the authors propose an approximation of the underlying state space by defining the space as a distance between the UE and the MEC server where the service is running. In this case, the time complexity is reduced to $O(N^2)$. As demonstrated by the numerical evaluations, the proposed migration strategy is able to decrease sum cost by roughly 35\% compared to both never and always migrate strategy. 

\textcolor{black}{The VM migration process may be further improved by a mobility prediction as demonstrated in \cite{Nademgega2016}. The proposed scheme is able to: 1) estimate in advance a throughput that user can receive from individual MEC servers as it roams throughout the network, 2) estimate time windows when the user perform handover, and 3) and VM migration management scheme selecting the optimal MEC servers according to offered throughput. The simulation results demonstrate that the proposed scheme is able to decrease latency by 35\% with respect to scheme proposed in \cite{Ksentini2014}. Nonetheless, a disadvantage of the proposal is that it requires huge amount of information in order to predict the throughput. Moreover, the paper does not consider the cost of migration itself.}   

In \cite{Wang2015c}, the VM migration decision process is further enhanced by the mechanism predicting future migration cost with specified upper bound on a prediction error. The main objective of the paper is, similarly as in \cite{Wang2014}\cite{Wang2015b}, to minimize the sum cost over a given time.  First, the authors propose an offline algorithm for finding the optimal placement sequence for a specific look-ahead window size $T$, which represents the time to which the cost prediction is done. For the offline algorithm, an arrival and a departure of the applications offloaded to the MEC are assumed to be exactly known. The time complexity of the algorithm is $O(M^2T)$, where $M$ stands for the number of MEC serves in the system. The VM migration is strongly dependent on the size of $T$. If $T$ is too large, the future predicted values may be far away from the actual values and, thus, the VM migration far from the optimal. Contrary if $T$ is too short, a long term effect of the service placement is not considered. As a result, also a binary search algorithm finding the optimal window size is proposed in the paper. The proposed offline algorithm is able to reduce cost by 25\% (compared to never migrate strategy) and by 32\% (compared to always migrate strategy). Although the simulation results are demonstrated for the multi-UEs scenario, the problem is formulated only for the single-UE. Hence, the paper is further extended in \cite{Wang2016b} for the multi-UEs offloading $K$ applications to the MEC. Similarly as in \cite{Wang2015c}, the problem is solved by the offline algorithm with complexity of $O(M^{K2}T)$. Since the offline algorithm is of high complexity and impractical for real systems, the paper also propose an online approximation algorithm reducing the complexity to $O(M^2KT)$. The proposed online algorithm outperforms never migrate and always migrate strategies by approximately 32\% and 50\%, respectively.

So far, all the studies focusing on the VM migration do not consider an impact on a workload scheduling, i.e., how the VM migration would be affected by a load of individual MEC servers. As suggested in \cite{Wang2015d}, the problem of the VM migration and scheduling of the MEC workloads should be done jointly. Although the problem could be formulated as a sequential decision making problem in the framework of MDPs (like in above studies) it would suffer from several drawbacks, such as, 1) extensive knowledge of the statistics of the users mobility and request arrival process is impractical, 2) problem can is computationally challenging, and 3) any change in the mobility and arrival statistics would require re-computing the optimal solution. Hence, the main contribution of \cite{Wang2015d} is a development of a new methodology overcoming these drawbacks inspired by Lyapunov optimization framework. The authors propose online control algorithm making decision on where the application should be migrated so that the overall transmission and reconfiguration costs are minimized. The complexity of the algorithm is $O(M!/(M-K)!)$, where $M$ is the number of MEC servers and $K$ is the amount of applications host by the MEC. By means of proposed optimization framework, the reconfiguration cost is reduced when compared to always migrate strategy (by 7\%) and never migrate strategy (by 26\%).

While the main objective of the previous papers focusing on the VM migration is to make a proper decision on whether to migrate or not, the main aim of the authors in \cite{Ha2015} is to minimize the VM migration time when the migration is about to be performed. This is accomplished by a compression algorithm reducing the amount of transmission data during the migration itself. On one hand, if the compression rate of the algorithm is low, more data has to be transmitted, but the compression itself is shorter in terms of time. On the other hand, a higher compression rate results in a significant reduction of the transmitted data during the VM migration, but the compression takes significant amount of time. Hence, the paper proposes a dynamic adaptation of the compression rate depending on the current backhaul available bandwidth and the processing load of the MEC. The paper presents extensive experiments on real system showing that the dynamic adaptation during the VM migration is able to cope with changing of available bandwidth capacity.

\textcolor{black}{A proper VM migration may not result only in an execution delay reduction, but it can also increase throughput of the system as demonstrated in \cite{Secci2016}. The paper proposes a protocol architecture for cloud access optimization (PACAO), which is based on Locator/Identifier Separation Protocol (LISP) \cite{LISP2013}. If the user is experiencing latency or jitter above maximum tolerated threshold, the VM migration to a new MEC server is initiated. The selection of the new MEC server, which is about to host VM of the user is based on the required computing power and availability of resources at the MEC servers. The proposal is evaluated by means of both experiments on real testbed and simulations. The results show that the system throughput is increased by up to 40\% when compared to the case without VM migration.}

\begin{figure}[b!]
	\centering
	\includegraphics[scale=0.11]{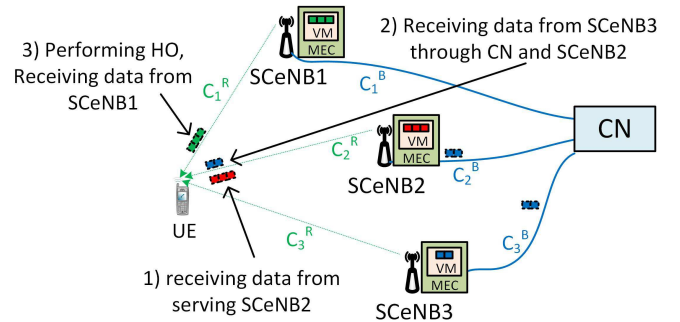} 
	\caption{An example of path selection algorithm proposed in \cite{Becvar2014} ($c_x^R$ and $C_x^B$ stands for capacity of radio links and backhaul links, respectively).}
	\label{fig:20}
\end{figure}

\begin{table*}[t!]
\footnotesize
\caption{The comparison of individual papers focusing on mobility management in MEC.}
\label{tab:Tab5}
\centering
\renewcommand{\arraystretch}{1}
\begin{tabular}{|p{0.40cm}|p{1.0cm}|p{2.8cm}|p{3.3cm}|p{1.8cm}|p{1.3cm}|p{2.8cm}|p{1.3cm}|}
\hline
	 & \bf Mobility man. method & \bf Objective & \bf Proposed method & \bf Mobility model & \bf Evaluation method & \bf Results & \bf Algorithm complexity \\ \hline
	\cite{Mach2014} & Power control & 1) Maximize the amount of delivered requests from the MEC, 2) Guaranteeing latency constraints  & Adaptation of transmission power of SCeNBs & 2D limited mobility (e.g.,  apartment) & Simulations & Up to \textbf{95\%} offloaded applications successfully delivered & - \\ \hline
	\cite{Mach2016} & Power control & 1) Maximize the amount of delivered requests from the MEC, 2) Guaranteeing latency constraints & Adaptation of transmission power of SCeNBs, optimization of power control trigger time & 2D limited mobility (e.g.,  apartment) & Simulations & Up to \textbf{98\%} offloaded applications successfully delivered & - \\ \hline
	\cite{Taleb2013b} & VM migration & 1) Define analytical model for VM migration, 2) Analyze how VM migration influences e2e delay  & - & 2D random walk model  & Analytical model & N/A & - \\ \hline
	\cite{Ksentini2014} & VM migration & 1) Maximize total expected reward  & Formulation of an optimal threshold decision policy exploiting MDP & 1D random walk & Simulations & Always maximize total expected reward wrt always/never migrate & - \\ \hline
\textcolor{black}{\cite{Sun2016} }&\textcolor{black}{VM migration} & \textcolor{black}{1) Find a trade-off between $Cost_M$ and $Gain_M$} & \textcolor{black}{Profit Maximization Avatar Placement  (PRIMAL) strategy deciding whether VM should be migrated or not} & \textcolor{black}{Random way point model} & \textcolor{black}{Simulations} & \textcolor{black}{Reducing execution delay by \textbf{90\%} wrt no migration, reducing migration cost by \textbf{40\%} wrt to always migrate} & \textcolor{black}{-} \\ \hline
	\cite{Wang2014} & VM migration & 1) Minimize system sum cost over a given time & Formulation of an optimal threshold decision policy using MDP & 1D asymmetric random walk mobility model& Simulations & Always minimize overall cost wrt always/never migrate & $O(|M|N)$ \\ \hline
		\cite{Wang2015b} & VM migration & 1) Minimize system sum cost over a given time & Formulation of an optimal sequential decision policy using MDP & 2D mobility, real mobility traces  & Simulations & \textbf{30\%} reduction of average cost wrt to never/always migrate & $O(N^2)$ \\ \hline
\textcolor{black}{\cite{Nademgega2016} }&\textcolor{black}{VM migration} & \textcolor{black}{1) Minimize execution delay} & \textcolor{black}{Estimation of throughput offered by MEC servers based on mobility prediction} & \textcolor{black}{Cars moving by a predefined paths} & \textcolor{black}{Simulations} & \textcolor{black}{Reducing latency by \textbf{35\%} wrt \cite{Ksentini2014}} & \textcolor{black}{-} \\ \hline
	\cite{Wang2015c} & VM migration & 1) Minimize system sum cost over a given time & Offline algorithm for finding optimal placement sequence for a specific look-ahead window size & Real world user mobility traces & Simulations & \textbf{25\%(32\%)} red. of average cost wrt to never(always) migrate & $O(M^2T)$ \\ \hline
	\cite{Wang2016b} & VM migration & 1) Minimize system sum cost over a given time & Offline and online algorithms for finding optimal placement sequence for a specific look-ahead window size & Real world user mobility traces & Analytical, simulations & \textbf{32\%(50\%)} red. of average cost wrt to never(always) migrate & $O(M^{K2}T)$ \\ \hline
	\cite{Wang2015d} & VM migration & 1) Minimize overall transmission and reconfiguration costs  & Online control algorithm making decision where application should be placed and migrated  & 1) Random walk, 2) Real world user mobility traces & Analytical evaluations, simulations & \textbf{7\%(26\%)} red. of reconfiguration cost wrt to never(always) migrate & $O(M!/(M -K)!)$ for $M \geq K$ \\ \hline
	\cite{Ha2015} & VM migration & 1) Minimize VM migration time & Adaptation of compression rate during VM migration depending on available bandwidth and processing load & - & Experiments on real system & N/A & - \\ \hline
\textcolor{black}{\cite{Secci2016} }&\textcolor{black}{ VM migration} & \textcolor{black}{1) Maximize throughput} & \textcolor{black}{Protocol architecture for cloud access optimization exploiting LISP} & \textcolor{black}{Real world user mobility traces} & \textcolor{black}{Experiments on testbed, simulations} & \textcolor{black}{Increase of throughput up to \textbf{40\%}} & \textcolor{black}{-} \\ \hline
	\cite{Becvar2014} & Path selection & 1) Minimize transmission delay & Path selection exploiting handover mechanism  & Manhattan mobility model & Simulations & Reduction of transmission delay by up to \textbf{9\%} & $O(m^n)$ \\ \hline
	\cite{Plachy2016} & Path selection & 1) Minimize transmission delay & Path selection exploiting handover mechanism & Manhattan mobility model & Simulations & Reduction of transmission delay by up to \textbf{54\%} & $O(Z^n)$ \\ \hline
	\cite{Plachy2016b} & Path selection + VM migration & 1) Minimize transmission delay & Cooperative service migraton and path selection algorithm with movement prediction & Smooth random mobility model & Simulations & Reduction of transmission delay by up to \textbf{10\%} wrt \cite{Plachy2016} & $O(|Z||I|\tau)$, $O(|I|\tau)$ \\ \hline
	\end{tabular}
\end{table*}

\subsection{Path selection and/or VM migration}
\label{sec53}
The VM migration is not a convenient option when a huge amount of data needs to be migrated among the computing nodes and the whole process may take minutes or even hours \cite{Ha2015}. Even if the migration process lasts few seconds, real-time applications cannot be offloaded to the MEC. Moreover, the load imposed on backhaul links may be too significant. In such cases, finding and optimizing new paths for delivery of the computed data from the MEC are a more viable option. This eventuality is considered in \cite{Becvar2014}, where the path selection algorithm for a delivery of the offloaded data from the cluster of computing SCeNBs to the UE is proposed. The main objective of the path selection algorithm is to minimize transmission delay taking into account quality of both radio and backhaul links. Moreover, the authors enable to enforce handover to new serving SCeNB to minimize the transmission delay. An example of the data delivery is shown in Fig.~\ref{fig:20}, where three SCeNBs are computing the application offloaded by the UE. The data processed by the serving SCeNB2 are received by the UE directly while data from the SCeNB3 are delivered to the UE through the CN and the serving SCeNB2. Finally, the UE performs handover to the SCeNB1 and, then, it receives results from the SCeNB3 directly via radio link. The complexity of the proposed algorithm is $O(m^n)$, where $m$ is the number of UEs and $n$ the amount of the SCeNBs in cluster. The proposed algorithm is able to reduce transmission delay by up to 9\% with respect to a case when the UE receives all data from the same serving SCeNB. In \cite{Plachy2016}, the algorithm's complexity is further decreased to $O(I^n)$, where $I$ is the set of SCeNBs with sufficient radio/backhaul link quality. It is shown that the proposed path selection algorithm is able to reduce transmission delay by 54\%.

The path selection algorithm contemplated in \cite{Becvar2014}\cite{Plachy2016} may not be sufficient if the UE is too far away from the computing location since increased transmission delay may result in QoS reduction notwithstanding. Hence, the authors in \cite{Plachy2016b} suggest a cooperation between an algorithm for the dynamic VM migration and the path selection algorithm proposed in \cite{Plachy2016} further enhanced by consideration of a mobility prediction. The first algorithm decides whether the VM migration should be initiated or not based on the mobility prediction and the computation/communication load of the eNB(s). The second algorithm, then, finds the most suitable route for downloading the offloaded data with the mobility prediction outcomes taken into account. The complexity of the first algorithm is $O(|Z||I|\tau)$ and the complexity of the second algorithm equals to $O(|I|\tau)$, where $Z$ is the number of eNBs with sufficient channel quality and computing capacity, and $\tau$ stands for the size of the prediction window. The proposed algorithm is able reducing the average offloading time by 27\% comparing to the situation when the VM migration is performed after each conventional handover and by roughly 10\% with respect to \cite{Plachy2016}.

\subsection{Summary of works focused on mobility management}
\label{sec54}
\textcolor{black}{A comparison of the studies addressing the mobility issues for the MEC is shown in Table~\ref{tab:Tab5}. As it can be observed from Table~\ref{tab:Tab5}, the majority of works so far focuses on the VM migration. Basically, the related papers try to find an optimal decision policy whether the VM migration should be initiated or not to minimize overall system cost (up to 32\% and up to 50\% reduction of average cost is achieved compared to never and always migrate options, respectively \cite{Wang2016b}). Moreover, some papers aim to find a proper trade-off between VM migration cost and VM migration gain \cite{Sun2016}, minimizing execution delay \cite{Nademgega2016}, minimizing VM migration time \cite{Ha2015}, or maximizing overall throughput \cite{Secci2016}.}

\textcolor{black}{From Table~\ref{tab:Tab5} can be further observed that all papers dealing with the VM migration assume the computation is done by a single computing node. Although this option is less complex, the parallel computation by more nodes should not be entirely neglected as most of the papers focusing on the allocation of computing resources assume multiple computing nodes (see Section~\ref{sec42}).}

\section{\textcolor{black}{Lessons learned}}
\label{LL}

\textcolor{black}{This section summarizes lessons learned from the state of the art focusing on computation offloading into the MEC. We again address all three key items: decision on computation offloading, allocation of computing resources, and mobility management.}

\textcolor{black}{From the surveyed papers dealing with the decision on computation offloading, following key observations are derived:
\begin{itemize}[leftmargin=12pt]
\item 
\textbf{If the channel quality between the UE and its serving station is low, it is profitable to compute rather locally} \cite{Munoz2015}. The main reason is that the energy spent by the transmission/reception of the offloaded data is too expensive in terms of the energy consumption at the UE. Contrary, \textbf{with increasing quality of the channel, it is better to delegate the computation to the MEC} since the energy cost required for transmission/reception of the offloaded data is reduced and it is easily outweighed by the energy saving due to the remote computation. Consequently, the \textbf{computation can be offloaded more frequently if MIMO is exploited} as it improves channel quality. Moreover, \textbf{it is efficient to exploit connection through SCeNBs for the offloading} as the SCeNBs are supposed to serve fewer users in proximity providing high channel quality and more available radio resources.
\item
\textbf{The most suitable applications for offloading are those requiring high computational power (i.e., high computational demanding applications) and, at the same time, sending only small amount of data} \cite{Sardellitti2014}. The reason is that the energy spent by transmission/reception of the offloaded computing is small while the energy savings achieved by the computation offloading are significant. Contrary, \textbf{the applications that need to offload a lot of data should be computed locally} as the offloading simply does not pay off due to huge amount of energy spent by the offloading and high offloading delays.  
\item
\textbf{If the computing capacities at the MEC are fairly limited, the probability to offload data for processing is lowered}. This is due to the fact that the probabilities of the offloading and local processing are closely related to the computation power available at the MEC.
\item
\textbf{With more UEs in the system, the application offloading as well as its processing at the MEC last longer} \cite{Munoz2014}. Consequently, if there is high amount of UEs in the system, the local processing may be more profitable, especially if the minimization of execution delay is the priority (such is the case of real-time applications).
\item
\textbf{The energy savings achieved by the computation offloading is strongly related to the radio access technology used at radio link.} To be more specific, OFDMA enables significantly higher energy savings of the UEs than TDMA due to higher granularity of radio resources \cite{You2016b}.
\item
\textbf{The partial offloading can save significantly more energy at the UE when compared to the full offloading} \cite{Deng2016}. Nevertheless, in order to perform the partial offloading, the application has to enable parallelization/partitioning. Hence, \textbf{the energy savings accomplished by computation offloading is also strongly related to the application type and the way how the code of the application is written}.
\end{itemize}
}

\textcolor{black}{
From the surveyed papers focused on allocation of computing resources, the following key facts are learned:
\begin{itemize}[leftmargin=12pt]
\item
\textbf{The allocation of computation resources is strongly related to the type of the application being offloaded} in a sense that only applications allowing parallelization/partitioning may be distributed to multiple computing nodes. Obviously, \textbf{a proper parallelization and code partitioning of the offloaded application can result in shorter execution delays} as multiple nodes may pool their computing resources (up to 90\% reduction of execution delay when compared to single computing node). On the other hand, \textbf{the allocation of computation resources for parallelized applications is significantly more complex}. 
\item
\textbf{An increase in the number of computing nodes does not have to result always in a reduction in the execution delay} \cite{Oueis2014}. On the contrary, if the communication delay becomes predominant over the computation delay, the overall execution delay may be even increased. Hence, a proper trade-off between the number of computing nodes and execution delay needs to be carefully considered when allocating computing resources to offloaded data.
\item
\textbf{If the backhaul is of a low quality, it is mostly preferred to perform the computation locally by the serving node} (e.g., SCeNB/eNB) since the distribution of data for computing is too costly in terms of the transmission latency. Contrary, a \textbf{high quality backhaul is a prerequisite for an efficient offloading to multiple computing nodes}.
\item
\textbf{The execution delay of the offloaded application depends not only on the backhaul quality, but also on a backhaul topology} (e.g., mesh, ring, tree, etc.) \cite{Oueis2014}. The mesh topology is the most advantageous in terms of the execution delay since all computing nodes are connected directly and distribution of the offloaded data for computing is more convenient. On the other hand, mesh topology would require huge investment in the backhaul.
\end{itemize} 
}

\textcolor{black}{
Finally, after surveying the papers addressing mobility issues in the MEC, we list following key findings:}

\begin{itemize}[leftmargin=12pt]
\item
\textcolor{black}{There are several options of the UE's mobility management if the data/application is offloaded to the MEC. \textbf{In cases of the low mobility, the power control at the SCeNBs/eNBs side can be sufficient to handle mobility} (up to 98\% of offloaded applications can be successfully delivered back to the UE \cite{Mach2016}). This is true as long as the adaption of transmission power enables keeping the UE at the same serving station during the computation offloading. However, if the UE performs handover, the power control alone is not sufficient and the VM migration or new communication path selection may be necessary to comply with requirements of offloaded applications in terms of latency.}
\textcolor{black}{
\item
A decision on VM migration depends strongly on three metrics: 
\begin{enumerate}[leftmargin=13pt]
\item
The \textbf{VM migration cost} ($Cost_M$) representing the time required for the service migration and the backhaul resources spent by the transmission of VM(s) between the computing nodes. 
\item
The \textbf{VM migration gain} ($Gain_M$) is the gain constituting delay reduction (data are computed in proximity of the UE) and saving of the backhaul resources (data does not have to be sent through several nodes). 
\item
The \textbf{computing load} of the node(s) to which the VM is reallocated since, in some situations, the optimal computing node for the VM migration may be unavailable due to its high computation load.
\end{enumerate}}

\item
\textcolor{black}{\textbf{The VM migration is impractical if huge amount of data needs to be transmitted between the computing nodes and/or if the backhaul resources between VMs are inadequate} since it may take minutes or even hours to migrate whole VM. This is obviously too long for real-time services and it also implies significant load on backhaul, especially if the VM migration would need to be performed frequently. Note that time consuming migration goes against the major benefit of the MEC, i.e., low latency resulting in suitability of the offloading for real-time services.
}
\item
\textcolor{black}{\textbf{The minimization of the VM migration time can be done by reduction of the amount of migrated data} \cite{Ha2015}. Nonetheless, even this option is not enough for real-time services. Thus, \textbf{various path selection algorithms should be employed with purpose to find the optimal path for delivery of the offloaded data back to the UEs} while computing is done by the same node(s) (i.e., without VM migration) \cite{Plachy2016}. However, \textbf{if the UE moves too far away from the computation placement, more robust mobility management based on joint VM migration and path selection should be adopted} \cite{Plachy2016b}.}
\end{itemize}

\section{Open research challenges and future work}
\label{sec6}
As shown in the previous sections, the MEC has attracted a lot of attention in recent years due to its ability to significantly reduce energy consumption of the UEs while, at the same time, enabling real-time application offloading because of proximity of computing resources to the users. Despite this fact the MEC is still rather immature technology and there are many challenges that need to be addressed before its implementation into mobile network to be beneficial. This section discusses several open research challenges not addressed by the current researcher.

\subsection{Distribution and management of MEC resources}
\label{sec61}
In Section~\ref{sec2}, we have discussed several possible options for placement of the computing nodes enabling the MEC within the mobile network architecture. To guarantee ubiquitous MEC services for all users wanting to utilize the MEC, the MEC servers and the computation/storage resource should be distributed throughout whole network. Consequently, the individual options where to physically place the MEC servers should complement each other in a hierarchical way. This will allow efficient usage of the computing resources while respecting QoS and QoE requirements of the users. In this context, an important challenge is to find an optimal way where to physically place the computation depending on expected users demands while, at the same time, consider related CAPEX and OPEX \textcolor{black}{(as initially tackled in \cite{Ceselli2015}\cite{Ceselli2017}).}

Another missing topic in the literature is a design of efficient control procedures for proper management of the MEC resources. This includes design of signalling messages, their exchange and optimization in terms of signalling overhead. The control messages should be able to deliver status information, such as load of individual computing nodes and quality of wireless/backhaul links in order to efficiently orchestrate computing resources within the MEC. There is a trade-off between high signalling overhead related to frequent exchange of the status information and an impact on the MEC performance due to aging of the status information if these are exchanged rarely. This trade-off have to be carefully analysed and efficient signalling mechanisms need to be proposed to ensure that the control entities in the MEC have up to date information at their disposal while the cost to obtain them is minimized.

\subsection{Offloading decision}
\label{sec62}
The offloading decision plays a crucial part as it basically determines whether the computation would be performed locally, remotely or jointly in both locations as discussed in Section~\ref{sec3}. All papers focusing on the offloading decision consider only the energy consumption at the side of the UE. However, to be in line with future green networking, also the energy consumption at the MEC (including computation as well as related communication) should be further taken into account during the decision. Moreover, all papers dealing with the offloading decision assume strictly static scenarios, i.e., the UEs are not moving before and during the offloading. Nevertheless, the energy necessary for transmission of the offloaded data can be significantly changed even during offloading if channel quality drops due to low movement or fading. This can result in the situation when the offloading may actually increase the energy consumption and/or execution delay comparing to local computation. Hence, it is necessary to propose new advanced methods for the offloading decision, for instance, exploiting various prediction techniques on the UEs mobility and channel quality during the offloading to better estimate how much the offloading will cost for varying conditions.  

Besides, current papers focusing on the partial offloading decision disregard the option to offload individual parts to multiple computing nodes. Multiple computing nodes enables higher flexibility and increases a probability that the offloading to the MEC will be efficient for the UE (in terms of both energy consumption and execution delay). Of course, a significant challenge in this scenario belongs to consideration of backhaul between the MEC servers and ability to reflect their varying load and parameters during the offloading decision.

\subsection{Allocation of computing resources}
\label{sec63}
The studies addressing the problem of an efficient allocation of the computing resources for the application offloaded to the MEC do not consider dynamicity of the network. To be more precise, the computing nodes (e.g., SCeNBs, eNB) are selected in advance before the application is offloaded to the MEC and then the same computing node(s) is (are) assumed to process the offloaded application (at least as long as the UE is relatively static and does not perform handover among cells as considered in Section~\ref{sec5}). However, if some additional computing resources are freed while given application is processed at the MEC, these resources could be also allocated for it in order to farther speed up the offloaded computing. Hence, a dynamic allocation of the computing resources during processing of the offloaded applications in the MEC is an interesting research challenge to be addressed in the future. 

So far all the studies focusing on the allocation of computing resources assume a "flat" MEC architecture in a sense that the MEC computing nodes are equally distributed and of the same computing power. In this respect, it would be interesting to consider more hierarchical placement of the computing nodes within the MEC. More specifically, computing resources should be distributed within the network as described in Section~\ref{sec223} (e.g., cluster of SCeNBs, eNBs, aggregation points or even at the edge of CN). A hierarchical MEC placement should result in a better distribution of the computing load and a lower execution delay experienced by the users since the use of distant CC can be more easily avoided. 

\subsection{Mobility management}
\label{sec64}
So far, the works focusing on mobility management and particularly on the VM migration consider mostly a scenario when only a single computing node (SCeNB or eNB) makes computation for each UE. Hence, the challenge is how to efficiently handle the VM migration procedure when application is offloaded to several computing nodes. Moreover, the VM migration impose high load on the backhaul and leads to high delay, which makes it unsuitable for real-time applications. Hence, new advanced techniques enabling very fast VM migration in order of milliseconds should be developed. However, this alternative is very challenging due to communication limits between computing nodes. Therefore, more realistic challenge is how to pre-migrate the computation in advance (e.g., based on some prediction techniques) so that there would be no service disruption observed by the users.

Despite of above-mentioned suggestions potentially reducing VM migration time, stand-alone VM migration may be unsuitable for real-time applications notwithstanding. Consequently, it is important to aim majority of research effort towards a cooperation of the individual techniques for mobility management. In this regard, dynamic optimization and joint consideration of all techniques (such as power control, VM migration, compression of migrated data, and/or path selection) should be studied more closely in order to enhance QoE for the UEs and to optimize overall system performance for moving users.

\subsection{Traffic paradigm imposed by coexistence of offloaded data and conventional data}
\label{sec65}
Current research dealing with the decision on computation offloading, allocation of computing resources and mobility management mostly neglects the fact that \textcolor{black}{conventional data not offloaded to the MEC}, such as VoIP, HTTP, FTP, machine type communication, video streaming, etc., has to be transmitted over radio and backhaul links in parallel to the offloaded data. \textcolor{black}{Hence, whenever any application is being offloaded to the MEC, it is necessary to jointly allocate/schedule communication resources both for the offloaded data to the MEC and the conventional data (i.e., data not exploiting MEC) in order to guarantee QoS and QoE.} Especially, if we consider the fact that the offloaded data represents additional load on already resource starving mobile cellular networks. The efficient scheduling of the communication resources may also increase the amount of data offloaded to the MEC because of more efficient utilization radio and backhaul communication links.
 
Besides, the offloading reshapes conventional perception of uplink/downlink utilization as the offloading is often more demanding in terms of the uplink transmission (offloading from the UE to the MEC). The reason for this is that many applications require delivering of large files/data to the MEC for processing (e.g., image/video/voice recognition, file scanning, etc.) while the results delivered to the UE are of significantly lower volume. This paradigm motivates for rethinking and reshaping research effort from sole downlink to the mixed downlink and uplink in the future.

\subsection{Concept validation}
\label{sec66}
As shown in Section~\ref{sec3}, ~\ref{sec4}, and ~\ref{sec5}, the MEC concept is analyzed and novel algorithms and proposed solutions are validated typically by numerical analysis or by simulations. In addition, majority of work assume rather simple, and sometimes unrealistic, scenarios for simplification of the problem. Although these are a good starting point in uncovering MEC potentials, it is important to validate key principles and findings by means of simulations under more complex and realistic situations and scenarios \textcolor{black}{such as, e.g., in \cite{Wang2015b}-\cite{Wang2015d} where at least real world user mobility traces are considered for evaluation and proposals on VM migration. At the same time, massive trials and further experiments in emulated networks (like initially provided in \cite{Dolezal2016}) or real networks (similar to those just recently performed by Nokia \cite{Nokia2016}) are mandatory to move the MEC concept closer to the reality.}

\section{Conclusion}
\label{sec7}
The MEC concept brings computation resources close to the UEs, i.e., to the edge of mobile network. This enables to offload highly demanding computations to the MEC in order to cope with stringent requirements of applications on latency (e.g., real time applications) and to reduce energy consumption at the UE. Although the research on the MEC gains its momentum, as reflected in this survey after all, the MEC itself is still immature and highly unproved technology. In this regard, the MEC paradigm introduces several critical challenges waiting to be addressed to the full satisfaction of all involved parties such as mobile operators, service providers, and users. The alpha and the omega of current research regarding the MEC is how to guarantee service continuity in highly dynamic scenarios. This part is lacking in terms of research and is one of the blocking point to enroll the MEC concept. Moreover, recent research validates solution mostly under very simplistic scenarios and by means of simulations or analytical evaluations. Nevertheless, to demonstrate the expected values introduced by the MEC, real tests and trials under more realistic assumptions are further required.

% or
%\appendix % for no appendix heading


%\section*{Acknowledgment}

%\section*{Acknowledgment}
%This work has been supported by Grant No. 13-24932P funded by the Czech Science Foundation

\ifCLASSOPTIONcaptionsoff
\newpage
\fi


\begin{thebibliography}{1}
\bibitem{Hoang2013}
Hoang T. Dinh, Chonho Lee, Dusit Niyato and Ping Wang, ``A survey of mobile cloud computing: architecture, applications, and approaches", \emph{Wireless Communications and Mobile Computing},  1587-1611, 13, 2013.

\bibitem{Barbarossa2014}
S. Barbarossa, S. Sardellitti, and P. Di Lorenzo, ``Communicating while Computing: Distributed mobile cloud computing over 5G heterogeneous networks", \emph{IEEE Signal Processing Magazine}, 31(6), 45-55, 2014.

\bibitem{Khan2014}
A. R. Khan, M. Othman, S. A. Madani, and S. U. Khan, ``A Survey of Mobile Cloud Computing
Application Models", \emph{IEEE Communications Surveys \& Tutorials}, 16(1), 393-413, First Quarter 2014.

\bibitem{Satyanarayanan2009}
M. Satyanarayanan, P. Bahl, R. Caceres, and N. Davies, ``The Case for VM-Based Cloudlets in Mobile Computing", \emph{IEEE Pervasive Computing}, 8(4), 14-23, 2009.

\bibitem{Shi2012}
C. Shi, V. Lakafosis, M. H. Ammar, and E. W. Zegura, ``Serendipity: Enabling remote computing among intermittently connected mobile devices", \emph{ACM international symposium on Mobile Ad Hoc Networking and Computing}, 145-154, 2012.

\bibitem{Drolia2013}
U. Drolia, et. al., ``The Case For Mobile Edge-Clouds," \emph{IEEE 10th International Conference on Ubiquitous Intelligence \& Computing and IEEE 10th International Conference on Autonomic \& Trusted Computing}, 209-215, 2013.

\bibitem{Mtibaa2013}
A. Mtibaa, A. Fahim, K. A. Harras, and M. H. Ammar, ``Towards resource sharing in mobile device clouds: Power balancing across mobile devices", \emph{ACM SIGCOMM workshop on Mobile cloud computing}, 51-56, 2013.

\bibitem{Mtibaa2013a}
A. Mtibaa, K. Harras, and A. Fahim, ``Towards computational offloading in mobile device clouds", \emph{IEEE International Conference on Cloud Computing Technology and Science}, 331-338, 2013.

\bibitem{Nishio2013}
T. Nishio, R. Shinkuma, T. Takahashi, and N. B. Mandayam, ``Service-oriented heterogeneous resource sharing for optimizing service latency in mobile cloud", \emph{Proceedings of the first international workshop on Mobile cloud computing \& networking}, 19-26, 2013.

\bibitem{Habak2015}
K. Habak, M. Ammar, K. A. Harras, and E. Zegura, ``Femto Clouds: Leveraging Mobile Devices to Provide Cloud Service at the Edge", \emph{IEEE International Conference on Cloud Computing}, 9-16, 2015.

\bibitem{Liu2013}
F Liu, P. Shu, H. Jin, L. Ding, J. Yu, D. Niu, and B. Li, ``Gearing resource-poor mobile devices with powerful clouds: architectures, challenges, and applications", \emph{IEEE Wireless Communications}, 20(3), 14-22, 2013.

\bibitem{Yaqoob2016}
I. Yaqoob, E. Ahmed, A. Gani, S. Mokhtar, M. Imran, and S. Guizani, ``Mobile ad hoc cloud: A survey", \emph{Wireless Communications and Mobile Computing}, 2016.

\bibitem{Ragona2015}
\textcolor{black}{C. Ragona, F. Granelli, C. Fiandrino, D. Kliazovich, and P. Bouvry, ``Energy-Efficient Computation Offloading for Wearable Devices and Smartphones in Mobile Cloud Computing", \emph{IEEE Global Communications Conference (GLOBECOM)}, 1-6, 2015.} 

\bibitem{Zhang2013a}
\textcolor{black}{W. Zhang, Y. Wen, J. Wu, and H. Li, ``Toward a unified elastic computing platform for smartphones with cloud support", \emph{IEEE Network}, 27(5), 34-40, 2013.} 

\bibitem{Bonomi2012}
F. Bonomi, R. Milito, J. Zhu, and S. Addepalli, ``Fog Computing and Its Role in the Internet of Things," \emph{MCC workshop on Mobile cloud computing}, 13-16, 2012.  

\bibitem{Zhu2013}
J. Zhu, D. S. Chan, M. S. Prabhu, P. Natarajan, H. Hu, and F. Bonomi, ``Improving Web Sites Performance Using Edge Servers in Fog Computing Architecture," \emph{IEEE International Symposium on Service-Oriented System Engineering}, 320-323, 2013. 

\bibitem{Stojmenovic2014}
I. Stojmenovic, S. Wen, ``The Fog Computing Paradigm: Scenarios and Security Issues," \emph{Federated Conference on Computer Science and Information Systems}, 1-8, 2014. 

\bibitem{Stojmenovic2014a}
I. Stojmenovic, ``Fog computing: A cloud to the ground support for smart things and machine-to-machine networks," \emph{Australasian Telecommunication Networks and Applications Conference (ATNAC)}, 117-122, 2014. 

\bibitem{Luan2015}
T. H. Luan, L. Gao, Y. Xiang, Z. Li, L. Sun, ``Fog Computing: Focusing on Mobile Users at the Edge. Available at https://arxiv.org/abs/1502.01815, 2016.

\bibitem{Yannuzzi2014}
M. Yannuzzi, R. Milito, R. Serral-Gracia, and D. Montero, ``Key ingredients in an IoT recipe: Fog Computing, Cloud computing, and more Fog Computing", \emph{International Workshop on Computer Aided Modeling and Design of Communication Links and Networks (CAMAD)}, 325-329, 2014.

\bibitem{Checko2015}
A. Checko, H. L. Christiansen, Y. Yan, L. Scolari, G. Kardaras, M. S. Berger, and L. Dittmann``Cloud RAN for Mobile Networks - A Technology Overview", \emph{IEEE Communications Surveys \& Tutorials}, 17(1), 405-426, First Quarter 2015. 

\bibitem{Kliazovich2008}
\textcolor{black}{D. Kliazovich and F. Granelli, ``Distributed Protocol Stacks: A Framework for Balancing Interoperability and Optimization", \emph{IEEE International Conference on Communications (ICC) Workshop}, 241-245, 2008.}

\bibitem{Cheng2016}
Ji. Cheng, Y. Shi, B. Bai, and W. Chen, ``Computation Offloading in Cloud-RAN Based
Mobile Cloud Computing System", \emph{IEEE International Conference on Communications (ICC)}, 1-6, 2016. 

\bibitem{Hu2015}
Y. Ch. Hu, M. Patel, D. Sabella, N. Sprecher, and V. Young, ``Mobile Edge Computing
A key technology towards 5G-First edition", 2015. 

\bibitem{Sanaei2014}
\textcolor{black}{Z. Sanaei, S. Abolfazli, A. Gani, and R. Buyya, ``Heterogeneity in Mobile Cloud Computing: Taxonomy and Open Challenges", \emph{IEEE Communications Surveys \& Tutorials}, 16(1), 369-392, First Quarter 2014.}

\bibitem{Wen2014}
\textcolor{black}{Y. Wen, X. Zhu, J. J. P. C. Rodrigues, and C. W. Chen, ``Cloud Mobile Media: Reflections and Outlook", \emph{IEEE Transactions on Multimedia}, 16(4), 885-902, June 2014.}

\bibitem{Ahmed2016}
A. Ahmed, E. Ahmed, ``A Survey on Mobile Edge Computing", \emph{IEEE International Conference on Intelligent Systems and Control (ISCO 2016)}, 1-8, 2016. 

\bibitem{Roman2016}
R. Roman, J. Lopez, M. Mambo, ``Mobile Edge Computing, Fog et al.: A Survey and Analysis of Security Threats and Challenges", \emph{Future Generation Computer Systems}, Nov 2016.

\bibitem{Luong2017}
\textcolor{black}{N. C. Luong, P. Wang, D. Niyato, W. Yonggang, and Z. Han, ``Resource Management in Cloud Networking Using Economic Analysis and Pricing Models: A Survey", \emph{IEEE Communications Surveys \& Tutorials}, PP(99), Jan. 2017.}

\bibitem{MAUI}
\textcolor{black}{E. Cuervo et al., ``MAUI: Making smartphones last longer with code offload," \emph{Int. Conf. Mobile Syst., Appl. Serv. (Mobysis)}, 49–62, 2010.}

\bibitem{Clonecloud}
\textcolor{black}{B. G. Chun, S. Ihm, P. Maniatis, M. Naik, and A. Patti, ``CloneCloud: Elastic execution between mobile device and cloud," \emph{Eur. Conf. Comput. Syst. (Eurosys)}, 301–314, 2011.}

\bibitem{ThinkAir}
\textcolor{black}{S. Kosta, A. Aucinas, P. Hui, R. Mortier, and X Zhang, ``ThinkAir: Dynamic resource allocation and parallel execution in the cloud for mobile code offloading", \emph{IEEE INFOCOM}, 945-953, 2012.}

\bibitem{Zhang2013b}
\textcolor{black}{W. Zhang, Y. Wen, and D. O. Wu, ``Energy-efficient Scheduling Policy for Collaborative Execution in Mobile Cloud Computing", \emph{IEEE INFOCOM}, 190-194, 2013.}

\bibitem{Zhang2015}
\textcolor{black}{W. Zhang, Y. Wen, and D. O. Wu, ``Collaborative task execution in mobile cloud computing under a stochastic wireless channel,” \emph{IEEE Trans. Wireless Commun.}, 14(1), 81–93, Jan. 2015.}

\bibitem{Wen2012}
\textcolor{black}{Y. Wen, W. Zhang, and H. Luo, ``Energy-Optimal Mobile Application Execution: Taming Resource-Poor Mobile Devices with Cloud Clones", \emph{IEEE INFOCOM}, 2716-20, 2012.}

\bibitem{Flores2015}
\textcolor{black}{H. Flores, P. Hui, S. Tarkoma, Y. Li, S. Srirama, and R. Buyya, ``Mobile Code Offloading: From Concept to Practice and Beyond", \emph{IEEE Communication Magazine}, 53(3), 80-88, 2015.}

\bibitem{Jiao2013}
\textcolor{black}{L. Jiao, R. Friedman, X. Fu, S. Secci, Z. Smoreda, and H. Tschofenig, ``Cloud-based Computation Offloading for Mobile Devices: State of the Art, Challenges and Opportunities", \emph{Future Network \& Mobile Summit}, 1-11, 2013.} 

\bibitem{MEC002}
ETSI GS MEC 002: Mobile Edge Computing (MEC); Technical Requirements V1.1.1, March 2016. 

\bibitem{Beck2014}
M. T. Beck, M. Werner, S. Feld, and T. Schimper, ``Mobile Edge Computing: A Taxonomy" \emph{International Conference on Advances in Future Internet (AFIN)}, 2014.

\bibitem{Taka2015}
\textcolor{black}{N. Takahashi, H. Tanaka, and R. Kawamura, ``Analysis of process assignment in multi-tier mobile cloud computing and application to Edge Accelerated Web Browsing", \emph{IEEE International Conference on Mobile Cloud Computing, Services, and Engineering}, 233-234, 2015.}

\bibitem{Zhang2012}
\textcolor{black}{Y. Zhang, H. Liu, L. Jiao, and X. Fu, ``To offload or not to offload: an efficient code partition algorithm for mobile cloud computing", \emph{1st International Conference on Cloud Networking (CLOUDNET)}, 80-86, 2012.}

\bibitem{Dolezal2016}
\textcolor{black}{J. Dolezal, Z. Becvar, and T. Zeman, ``Performance Evaluation of Computation Offloading from Mobile Device to the Edge of Mobile Network", \emph{IEEE Conference on Standards for Communications and Networking (CSCN)}, 1-7, 2016.}

\bibitem{Salman2015}
\textcolor{black}{O. Salman, I. Elhajj, A. Kayssi, and A. Chehab, ``Edge Computing Enabling the Internet of Things", \emph{IEEE 2nd World Forum on Internet of Things (WF-IoT)}, 603-608, 2015.}

\bibitem{Ab2016}
\textcolor{black}{S. Abdelwahab, B. Hamdaoui, M. Guizani, and T. Znati, ``REPLISOM: Disciplined Tiny Memory Replication for Massive IoT Devices in LTE Edge Cloud", \emph{IEEE Internet of Things Journal}, 3(3), 327-338, 2016.}

\bibitem{Sun2016a}
\textcolor{black}{X. Sun and N. Ansari, ``EdgeIoT: Mobile Edge Computing for the Internet of Things", \emph{IEEE Communications Magazine}, 54(12), 22-29, 2016.}

\bibitem{Nokia2016}
\textcolor{black}{NOKIA: Multi-access Edge Computing, [online] https://networks. nokia.com/solutions/multi-access-edge-computing.}

\bibitem{Nokia2016a}
\textcolor{black}{NOKIA: Mobile Edge Computing, [online] http://resources.alcatel-lucent.com/asset/200546.}

\bibitem{TROPIC2012}
FP7 European Project, Distributed computing, storage and radio
resource allocation over cooperative femtocells (TROPIC). [Online]. Available:
http://www.ict-tropic.eu/, 2012. 

\bibitem{SESAM2015}
H2020 European Project, Small cEllS coordinAtion for Multi-tenancy and Edge services (SESAM), Available: http://www.sesame-h2020-5g-ppp.eu/, 2015. 

\bibitem{Giannoulakis2016}
I. Giannoulakis, et. al., ``The emergence of operator-neutral small cells as a strong case for cloud computing at the mobile edge", \emph{Trans. Emerging Tel. Tech.}, 27, 1152–59, 2016. 

\bibitem{NFV2012}
M. Chiosi, et. al. ``Network Functions Virtualisation: An Introduction, Benefits, Enablers, Challenges \& Call for Action", Introductory white paper, 2012. 

\bibitem{NFV2013}
ETSI GS NFV 002: Architectural Framework, V1.1.1, Oct. 2013. 

\bibitem{Lobillo2014}
F. Lobillo, et. al., ``An architecture for mobile computation offloading on cloud-enabled LTE small cells", \emph{Workshop on Cloud Technologies and Energy Efficiency in Mobile Communication Networks (IEEE WCNCW)}, 1-6, 2014. 

\bibitem{Puente2015}
M. A. Puente, Z. Becvar, M. Rohlik, F. Lobillo, and E. C. Strinati, ``A Seamless Integration of
Computationally-Enhanced Base Stations into Mobile Networks towards 5G", \emph{IEEE Vehicular Technology Conference (IEEE VTC-Spring 2015) workshops}, 1-5, 2015. 

\bibitem{Becvar2017}
Z. Becvar, M. Rohlik, P. Mach, M. Vondra, T. Vanek, M.A. Puente, F. Lobillo, ``Distributed Architecture of 5G Mobile Networks for Efficient Computation Management in Mobile Edge Computing", \emph{Chapter in 5G Radio Access Network (RAN) - Centralized RAN, Cloud-RAN and Virtualization of Small Cells (H. Venkataraman, R. Trestian, eds.)}, Taylor and Francis Group, USA, March 2017. 

\bibitem{Wang2013}
S. Wang, et. al., ``Mobile Micro-Cloud: Application Classification, Mapping, and Deployment", \emph{Annual Fall Meeting of ITA (AMITA)}, 2013. 

\bibitem{Wang2015}
K. Wang, M. Shen, and J. Cho, ``MobiScud: A Fast Moving Personal Cloud in the Mobile Network", \emph{Workshop on All Things Cellular: Operations, Applications and Challenge}, 19-24, 2015. 

\bibitem{Manzalini2014}
A. Manzalini, et. al., ``Towards 5G Software-Defined Ecosystems: Technical Challenges, Business Sustainability and Policy Issues", white paper, 2014.

\bibitem{Taleb2013}
T. Taleb, A. Ksentini, ``Follow Me Cloud: Interworking Federated Clouds and Distributed Mobile Networks", \emph{IEEE Network}, 27(5), 12-19, 2013. 

\bibitem{Taleb2016}
T. Taleb, A. Ksentini, and P. A. Frangoudis, ``Follow-Me Cloud: When Cloud Services Follow
Mobile Users", \emph{IEEE Transactions on Cloud Computing}, PP(99), 1-1, 2016. 

\bibitem{Aissioui2015}
A. Aissioui, A. Ksentini, and A. Gueroui, ``An Efficient Elastic Distributed SDN Controller for
Follow-Me Cloud", \emph{IEEE International Conference on Wireless and Mobile Computing, Networking and Communications (WiMob)}, 876-881, 2015. 

\bibitem{Liu2014}
J. Liu, T. Zhao, S. Zhou, Y. Cheng, and Z. Niu, ``CONCERT: a cloud-based architecture for next-generation cellular systems", \emph{IEEE Wireless Communications}, 21(6), 14-22, Dec. 2014. 

\bibitem{MEC001}
ETSI GS MEC 001: Mobile Edge Computing (MEC); Terminology V1.1.1, March 2016. 

\bibitem{MEC005}
ETSI GS MEC 005: Mobile Edge Computing (MEC); Proof of Concept Framework V1.1.1, March 2016. 

\bibitem{MEC004}
ETSI GS MEC 004: Mobile Edge Computing (MEC); Service Scenarios V1.1.1, March 2016. 

\bibitem{MEC003}
ETSI GS MEC 003: Mobile Edge Computing (MEC); Framework and Reference Architecture V1.1.1, March 2016. 

\bibitem{Ceselli2015}
\textcolor{black}{A. Ceselli, M. Premoli, and S. Secci, ``Cloudlet Network Design Optimization", \emph{IFIP Networking}, 1-9, 2015.}

\bibitem{Ceselli2017}
A. Ceselli, M. Premoli, and S. Secci, ``Mobile Edge Cloud Network Design Optimization", \emph{IEEE/ACM Transactions on Networking}, PP(99), 2017.

\bibitem{Kreutz2015}
\textcolor{black}{D. Kreutz, F. M. V. Ramos, P. E. Verissimo, C. E. Rothenberg, S. Azodolmolky, and S. Uhlig, ``Software-Defined Networking: A Comprehensive Survey", \emph{Proceedings of the IEEE}, 103(1), 14-76, Jan. 2015.}

\bibitem{Jaga2015}
\textcolor{black}{N. A. Jagadeesan and B. Krishnamachari, ``Software-Defined Networking Paradigms in Wireless Networks: A Survey", \emph{ACM Computing Surveys}, 47(2), Jan. 2015.}

\bibitem{Jin2013}
\textcolor{black}{X. Jin, L. E. Li, L. Vanbever, and J. Rexford, ``SoftCell: scalable and flexible cellular core network architecture", \emph{ACM conference on Emerging networking experiments and technologies (CoNEXT)}, 163-174, 2013.}

\bibitem{Deng2016}
M. Deng, H. Tian, and B. Fan, ``Fine-granularity Based Application Offloading Policy in Small Cell Cloud-enhanced Networks", \emph{IEEE International Conference on Communications Workshops (ICC)}, 638-643, 2016.

\bibitem{Mahmoodi2016} 
\textcolor{black}{S. E. Mahmoodi, R. N. Uma, and K. P. Subbalakshmi, ``Optimal joint scheduling and cloud offloading for mobile applications," \emph{IEEE Trans. Cloud Comput.}, PP(99), 2016.}

\bibitem{Liu2016}
J. Liu, Y. Mao, J. Zhang, and K. B. Letaief, ``Delay-Optimal Computation Task Scheduling for
Mobile-Edge Computing Systems", \emph{IEEE International Symposium on Information Theory (ISIT)}, 1451-55, 2016.  

\bibitem{Mao2016}
Y. Mao, J. Zhang, K. B. Letaief, ``Dynamic Computation Offloading for Mobile-Edge Computing with Energy Harvesting Devices",\emph{ IEEE Journal on Selected Areas in Communications}, 34(12), 3590-3605, 2016. 

\bibitem{Zhang2013}
W. Zhang, Y. Wen, K. Guan, D. Kilper, H. Luo, and D. O. Wu, ``Energy-Optimal Mobile Cloud Computing under Stochastic Wireless Channel", \emph{IEEE Transactions on Wireless Communications}, 12(9), 4569-81, 2013. 

\bibitem{Ulukus2015}
S. Ulukus, A. Yener, E. Erkip, O. Simeone, M. Zorzi, P. Grover, and K. Huang, ``Energy Harvesting Wireless Communications: A Review of Recent Advances", \emph{IEEE Journal on Selected Areas in Communications}, 33(3), 360-381, 2015. 

\bibitem{Kamoun2015}
M. Kamoun, W. Labidi, and M. Sarkiss, ``Joint resource allocation and offloading strategies
in cloud enabled cellular networks",\emph{ IEEE International Conference on Communications (ICC)}, 5529-34, 2015. 

\bibitem{Labidi2015}
W. Labidi, M. Sarkiss, and M. Kamoun, ``Energy-Optimal Resource Scheduling and
Computation Offloading in Small Cell Networks", \emph{International Conference on Telecommunications (ICT)}, 313-318, 2015. 

\bibitem{Labidi2015b}
W. Labidi, M. Sarkiss, and M. Kamoun, ``Joint Multi-user Resource Scheduling and Computation Offloading in Small Cell Networks", \emph{IEEE International Conference onWireless and Mobile Computing, Networking and Communications (WiMob)}, 794-801, 2015. 

\bibitem{Barbarossa2013}
S. Barbarossa, S. Sardellitti, and P. Di Lorenzo, ``Joint allocation of computation and communication resources in multiuser mobile cloud computing", \emph{IEEE Workshop on Signal Processing Advances in Wireless Communications (SPAWC)}, 26-30, 2013. 

\bibitem{Sardellitti2014}
S. Sardellitti, G. Scutari, and S. Barbarossa, ``Joint Optimization of Radio and Computational
Resources for Multicell Mobile Cloud Computing", \emph{IEEE International Workshop on Signal Processing Advances in Wireless Communications (SPAWC)}, 354-358, 2014. 

\bibitem{Sardellitti2014b}
S. Sardellitti, S. Barbarossa, G. Scutari, ``Distributed Mobile Cloud Computing: Joint Optimization of Radio and Computational Resources", \emph{IEEE Globecom Workshops (GC Wkshps)}, 1505-10, 2014. 

\bibitem{Zhang2016}
K. Zhang, Y. Mao, S. Leng, Q. Zhao, L. Li, X. Peng, L. Pan, S. Maharjan, and Y. Zhang, ``Energy-Efficient Offloading for Mobile Edge Computing in 5G Heterogeneous Networks", \emph{IEEE Access}, 4, 5896-5907, 2016. 

\bibitem{Chen2016}
X. Chen, L. Jiao, W. Li, and X. Fu, ``Efficient Multi-User Computation Offloading for
Mobile-Edge Cloud Computing", \emph{IEEE/ACM Transactions on Networking}, 24(5), 2795-2808, 2016. 

\bibitem{Chen2015}
M.-H. Chen, B. Liang, and M. Dong, ``A Semidefinite Relaxation Approach to Mobile
Cloud Offloading with Computing Access Point", \emph{IEEE International Workshop on Signal Processing Advances in Wireless Communications (SPAWC)}, 186-190, 2015. 

\bibitem{Chen2016b}
M.-H. Chen, M. Dong, and B. Liang, ``Joint offloading decision and resource allocation for mobile cloud with computing access point", \emph{IEEE International Conference on Acoustics, Speech and Signal Processing (ICASSP)}, 3516-20, 2016. 

\bibitem{Cao2015}
S. Cao, X. Tao,Y. Hou, and Q. Cui, ``An Energy-Optimal Offloading Algorithm of Mobile Computing Based on HetNets", \emph{International Conference on Connected Vehicles and Expo (ICCVE)}, 254-258, 2015. 

\bibitem{Kennedy1997}
J. Kennedy and R. C. Eberhart, ``A Discrete Binary Version of the Particle Swarm Algorithm", \emph{IEEE International Conference on Systems, Man, and Cybernetics}, 4104-08, 1997. 

\bibitem{Zhao2015}
Y. Zhao, S. Zhou, T. Zhao, and Z. Niu, ``Energy-Efficient Task Offloading for Multiuser Mobile Cloud Computing", \emph{IEEE/CIC International Conference on Communications in China (ICCC)}, 1-5, 2015. 

\bibitem{You2016}
C. You and K. Huang, ``Multiuser Resource Allocation for Mobile-Edge Computation Offloading", \emph{IEEE Global Communication Conference (GLOBECOM)}, 1-6, 2016. 

\bibitem{You2016b}
C. You, K. Huang, H. Chae, and B.-H. Kim, ``Energy-Efficient Resource Allocation for Mobile-Edge Computation Offloading", \emph{IEEE Transactions on Wireless Communications}, 16(3), 1397-1411, 2017. 

\bibitem{Wang2016c}
\textcolor{black}{Y. Wang, M. Sheng, X. Wang, L. Wang, and J. Li, ``Mobile-Edge Computing: Partial Computation Offloading Using Dynamic Voltage Scaling", \emph{IEEE Transactions on Communications}, 64(10), pp. 4268-82, 2016.}

\bibitem{Munoz2013}
O. Munoz, A. Pascual-Iserte, and J. Vidal, ``Joint Allocation of Radio and Computational Resources in Wireless Application Offloading", \emph{Future Network and Mobile Summit}, 1-10, 2013. 

\bibitem{Munoz2015}
O. Munoz, A. Pascual-Iserte, and J. Vidal, ``Optimization of Radio and Computational Resources
for Energy Efficiency in Latency-Constrained Application Offloading", \emph{IEEE Transactions on Vehicular Technology}, 64(10), 4738-55, 2015. 

\bibitem{Munoz2014}
O. Munoz, A. Pascual-Iserte, J. Vidal, and M. Molina, ``Energy-Latency Trade-off for Multiuser Wireless Computation Offloading", \emph{IEEE Wireless Communications and Networking Conference Workshops (WCNCW)}, 29-33, 2014. 

\bibitem{Mao2016b}
Y. Mao, J. Zhang, S.H. Song, and K. B. Letaief, ``Power-Delay Tradeoff in Multi-User Mobile-Edge Computing Systems",  \emph{IEEE Global Communications Conference (GLOBECOM)}, 1-6, Dec. 2016.

\bibitem{Zhao2015b}
T. Zhao, S. Zhou, X. Guo, Y. Zhao, and Z. Niu, ``A Cooperative Scheduling Scheme of Local Cloud and Internet Cloud for Delay-Aware Mobile Cloud Computing", \emph{IEEE Globecom Workshops (GC Wkshps)}, 1-6, 2015.

\bibitem{Guo2016}
X. Guo, R. Singh, T. Zhao, and Z. Niu, ``An Index Based Task Assignment Policy for Achieving Optimal Power-Delay Tradeoff in Edge Cloud Systems", \emph{IEEE International Conference on Communications (ICC)}, 1-7, 2016.

\bibitem{Valerio2014}
V. Di Valerio and F. Lo Presti, ``Optimal Virtual Machines Allocation in Mobile Femto-cloud Computing: an MDP Approach", \emph{IEEE Wireless Communications and Networking Conference Workshops (WCNCW)}, 7-11, 2014. 

\bibitem{Tanzil2015}
S. M. S. Tanzil, O. N. Gharehshiran, and V. Krishnamurthy, ``Femto-Cloud Formation: A Coalitional Game-Theoretic Approach", \emph{IEEE Global Communications Conference (GLOBECOM)}, 1-6, 2015.

\bibitem{Oueis2014}
J. Oueis, E. Calvanese-Strinati, A. De Domenico, and S. Barbarossa, ``On the Impact of Backhaul Network on Distributed Cloud Computing", \emph{IEEE Wireless Communications and Networking Conference Workshops (WCNCW)}, 12-17, 2014.

\bibitem{Oueis2014b}
J. Oueis, E. Calvanese Strinati, and S. Barbarossa, ``Small Cell Clustering for Efficient Distributed Cloud Computing", \emph{IEEE Annual International Symposium on Personal, Indoor, and Mobile Radio Communication (PIMRC)}, 1474-79, 2014.

\bibitem{Oueis2015}
J. Oueis, E. Calvanese Strinati, S. Sardellitti, and S. Barbarossa, ``Small Cell Clustering for Efficient Distributed Fog Computing: A Multi-User Case", \emph{IEEE Vehicular Technology Conference (VTC Fall)}, 1-5, 2015.

\bibitem{Oueis2015b}
J. Oueis, E. Calvanese Strinati, and S. Barbarossa, ``The Fog Balancing: Load Distribution for Small Cell Cloud Computing", \emph{IEEE 81st Vehicular Technology Conference (VTC Spring)} 1-6, 2015.

\bibitem{Vondra2014}
M. Vondra and Z. Becvar, ``QoS-ensuring Distribution of Computation Load among Cloud-enabled Small Cells", \emph{IEEE International Conference on Cloud Networking (CloudNet)}, 197-203, 2014. 

\bibitem{Wang2016}
S. Wang, M. Zafer, and K. K. Leung, ``Online Placement of Multi-Component Applications in Edge Computing Environments", \emph{IEEE Access}, PP(99), 2017.

\bibitem{Mach2014}
P. Mach and Z. Becvar, ``Cloud-aware power control for cloud-enabled small cells", \emph{IEEE Globecom Workshops (GC Wkshps)}, 1038-43, 2014.

\bibitem{Mach2016}
P. Mach and Z. Becvar, ``Cloud-aware power control for real-time application offloading in mobile edge computing", \emph{Transactions on Emerging Telecommunications Technologies}, 2016.

\bibitem{Taleb2013b}
T. Taleb and A. Ksentini, ``An Analytical Model for Follow Me Cloud", \emph{IEEE Global Communications Conference (GLOBECOM)}, 1291-96, 2013.

\bibitem{Ksentini2014}
A. Ksentini, T. Taleb, and M. Chen, ``A Markov Decision Process-based Service Migration Procedure for Follow Me Cloud", \emph{IEEE International Conference on Communications (ICC)}, 1350-54, 2014

\bibitem{Sun2016}
\textcolor{black}{X. Sun and N. Ansari, ``PRIMAL: PRofIt Maximization Avatar pLacement for Mobile Edge Computing", \emph{IEEE International Conference on Communications (ICC)}, 1-6, 2016.}

\bibitem{Wang2014}
S. Wang, R. Urgaonkar, T. He, M. Zafer, K. Chan, and K. K. Leung, ``Mobility-Induced Service Migration in Mobile Micro-Clouds", \emph{IEEE Military Communications Conference}, 835-840, 2014.

\bibitem{Wang2015b}
S. Wang, R. Urgaonkar, M. Zafer, T. He, K. Chan, and K. K. Leung, ``Dynamic Service Migration in Mobile Edge-Clouds", \emph{IFIP Networking Conference (IFIP Networking)}, 1-9, 2015.

\bibitem{Nademgega2016}
\textcolor{black}{A. Nadembega, A. S. Hafid, and R. Brisebois, ``Mobility Prediction Model-based Service Migration Procedure for Follow Me Cloud to support QoS and QoE", \emph{IEEE International Conference on Communications (ICC)}, 1-6, 2016.}

\bibitem{Wang2015c}
S. Wang, R. Urgaonkar, K. Chan, T. He, M. Zafer, and K. K. Leung, ``Dynamic Service Placement for Mobile Micro-Clouds with Predicted Future Costs", \emph{IEEE International Conference on Communications (ICC)}, 5504-10, 2015.

\bibitem{Wang2016b}
S. Wang, R. Urgaonkar, T. He, K. Chan, M. Zafer, and K. K. Leung, ``Dynamic Service Placement for Mobile Micro-Clouds with Predicted Future Costs", \emph{IEEE Transactions on Parallel and Distributed Systems}, PP(99), 2016.

\bibitem{Wang2015d}
R. Urgaonkar, S. Wang, T. He, M. Zafer, K. Chan, and K. K. Leung, ``Dynamic service migration and workload scheduling in edge-clouds", \emph{Performance Evaluations}, 91(2015), 205-228, 2015.

\bibitem{Ha2015}
K. Ha, Y. Abe, Z. Chen, W. Hu, B. Amos, P. Pillai, and M. Satyanarayanan, ``Adaptive VM Handoff Across Cloudlets", Technical Report CMU-CS-15-113, Computer Science Department,  June 2015.

\bibitem{Secci2016}
\textcolor{black}{S. Secci, P. Raad, and P. Gallard, ``Linking Virtual Machine Mobility to User Mobility", \emph{IEEE Transactions on Network and Service Management}, 13(4), pp. 927-940, 2016.}

\bibitem{LISP2013}
\textcolor{black}{D. Farinacci, V. Fuller, D. Meyer, and D. Lewis, ``The locator/ID separation
protocol (LISP)," \emph{Internet Eng. Task Force}, Fremont, CA, USA,
RFC 6830, 2013.}

\bibitem{Becvar2014}
Z. Becvar, J. Plachy, and P. Mach, ``Path Selection Using Handover in Mobile Networks with Cloud-enabled Small Cells", \emph{IEEE International Symposium on Personal, Indoor and Mobile Radio Communications (PIMRC)}, 1480-85, 2014.

\bibitem{Plachy2016}
J. Plachy, Z. Becvar, and P. Mach, ``Path Selection Enabling User Mobility and Efficient Distribution of Data for Computation at the Edge of Mobile Network", \emph{Computer Networks}, 108, 357-370, 2016.

\bibitem{Plachy2016b}
J. Plachy, Z. Becvar, and E. Calvanese Strinati, ``Dynamic Resource Allocation Exploiting Mobility Prediction in Mobile Edge Computing", \emph{IEEE International Symposium on Personal, Indoor and Mobile Radio Communications (PIMRC)}, 1-6, 2016.

\end{thebibliography}
\end{document}